\documentclass[pre,floatfix,twocolumn,a4paper,nofootinbib,amssymb,amsmath,superscriptaddress]{revtex4}
\usepackage{graphicx}
\usepackage{bm}
\usepackage{eucal}% altera il mathcal
\usepackage{mathrsfs}% per i caratteri col \mathscr
\usepackage{amsmath}
\usepackage{color,colortbl}
\usepackage{soul}

\begin{document}

\title{Neural Activity of Heterogeneous Inhibitory Spiking Networks with Delay}

\author{Stefano Luccioli}
\affiliation{CNR - Consiglio Nazionale delle Ricerche - Istituto dei Sistemi Complessi, via Madonna del Piano 10, 50019 Sesto Fiorentino, Italy}
\author{David Angulo Garcia}
\affiliation{Grupo de Modelado Computacional - Din\'{a}mica y Complejidad de Sistemas. Instituto de Matem\'{a}ticas Aplicadas. Universidad de Cartagena. Calle de la Universidad \# 6 - 100, Cartagena de Indias, Colombia}
\author{Alessandro Torcini}
\affiliation{Laboratoire de Physique Th\'eorique et Mod\'elisation, Universit\'e de Cergy-Pontoise, CNRS, UMR 8089,
95302 Cergy-Pontoise cedex, France}
\affiliation{CNR - Consiglio Nazionale delle Ricerche - Istituto dei Sistemi Complessi, via Madonna del Piano 10, 50019 Sesto Fiorentino, Italy}

\date{\today}

\begin{abstract}
We study a network of spiking neurons with heterogeneous excitabilities
connected via inhibitory delayed pulses. For globally coupled systems 
the increase of the inhibitory coupling reduces the number of firing neurons 
by following a Winner Takes All mechanism.
For sufficiently large transmission delay we observe the emergence
of collective oscillations in the system beyond a critical coupling value. 
Heterogeneity promotes neural inactivation and asynchronous dynamics and
its effect can be counteracted by considering longer time delays.
In sparse networks,  inhibition has the counterintuitive effect of promoting neural 
reactivation of silent neurons for sufficiently large coupling. In this regime, 
current fluctuations are on one side responsible for neural firing
of sub-threshold neurons and on the other side for their desynchronization.
Therefore, collective oscillations are present only in a limited range of
coupling values, which remains finite in the thermodynamic limit.
Out of this range the dynamics is asynchronous and for very large inhibition 
neurons display a bursting behaviour alternating
periods of silence with periods where they fire freely in absence of any inhibition.
\end{abstract}

\maketitle

\section{Introduction}

Despite the fact that inhibition emerges only
at later stages of development of the brain\cite{ben2001developing},
its role is fundamental for a correct and healthy functioning 
of the cerebral circuits.
In the adult brain the majority of neurons are excitatory, while only 15-20 \%
has been identified as inhibitory interneurons. However
this limited presence  is sufficient to
allow for an overall homeostatic regulation of global 
activity in the cerebral cortex and at the same time
for rapid changes in local excitability, which are needed
to modify network connections and for processing information \cite{jonas2007}.

The role of inhibition in promoting brain rhythms at a mesoscopic level,
in particular in the beta (12-30 Hz) and gamma (30-100 Hz) bands,
has been clearly demonstrated in experiments and
network models \cite{whittington2000, Buzsaki2006Rhythms}.
Furthermore, inhibition induced oscillations provide 
a temporal framing for the discharge patterns of excitatory cells 
possibly associated to locomotory behaviours or cognitive functions
\cite{buzsaki1995,salinas2001}.

This justifies the interest for studying the dynamics of purely 
inhibitory neural networks and in particular for the emergence
of collective oscillations (COs) in these systems. COs have been usually reported for networks 
presenting either a time delay in the transmission of the 
neural signal or a finite synaptic time scale.
An interesting analogy can be traced between 
the dynamics of inhibitory networks with delay and instantaneous synapses and that of circuits where the
post-synaptic potential (PSP) has a finite duration, but with 
no delay in the synaptic transmission.
In particular, in \cite{vanVreeswijk1996Partial} it has been shown that
in homogeneous fully coupled networks for finite PSPs one usually observes 
coexistence of synchronized clusters of different sizes, analogously to what
reported in \cite{ernst1995,kinzel2009} for delayed systems. Furthermore, in
\cite{ernst1995} the authors found that the average number of coexisting clusters
decreases with the delay, somehow analogously to what
reported in \cite{vanVreeswijk1996Partial} for increasing duration of the PSP.
As a matter of fact stable splay states (corresponding to a number of clusters
equal to the number of neurons) are observable in the limit of
zero delay and instantaneous synapses \cite{Zillmer2006}.
 
The introduction of disorder in the network, at the level of
connection or excitability distributions, does not prevent the emergence 
of COs, as shown for systems with delay \cite{BrunelHakim1999,Brunel2000Sparse,politi2010lif}
or with finite PSPs \cite{golomb2000number}.
The only case in which COs have been reported in sparse networks in absence of 
delay and for instantaneous synapses is for Quadratic Integrate-and-Fire (QIF) neurons
in a balanced regime \cite{di2018}.

A common phenomenon observable in inhibitory networks
is the progressive silencing ({\it neurons' death}) of less
excitable neurons induced by the activity of the most excitable ones 
when the inhibition increase. This mechanism,  
referred in the literature as Winner  Takes  All (WTA)
with inhibitory feedback \cite{coultrip1992cortical, fukai1997simple,itti2001},
has been employed to explain
 attention activate competition among visual filters \cite{lee1999}, 
visual discrimination tasks \cite{wang2002probabilistic,wong2006recurrent},
as well as the so-called $\gamma$-cycle documented in several brain regions~\cite{fries2007}.
Furthermore, the WTA mechanism has been demonstrated to emerge in inhibitory
spiking networks for heterogeneous 
distributions of the neural excitabilities \cite{angulo2017death_NJP}.
However, while in globally coupled networks (GCNs) the increase in synaptic inhibition 
can finally lead to only few or even only one surviving neuron,
in sparse networks (SNs) inhibition can astonishingly promote, 
rather than depress, neural activity inducing the reactivation of silent neurons
\cite{ponzi2013optimal, angulo2015Striatum, angulo2017death_NJP}.

Our aim is to analize in neural networks with delay the combined effect of synaptic inhibition
and different types of disorder on neurons' death and reactivation, as well as on the emergence of COs.
In particular, we will first investigate GCNs showing that in this 
case despite the number of active neurons steadily decrease with the
inhibition, due to the WTA mechanism, COs can emerge for sufficiently large
synaptic inhibition and delay. An increase of the heterogeneity in the neural 
excitabilities promotes neural's death and asynchronous behaviour in the network.
This effect can be somehow compensated by considering longer time delays.

In SNs, at sufficiently large synaptic coupling current fluctuations, induced by the disorder in the pre-synaptic connections \cite{BrunelHakim1999,Brunel2000Sparse}, are responsible for the firing of
inactive neurons. At the same time, these fluctuations desynchronize the 
neural activity leading to the disappearence of COs. 
Therefore in SNs by varying the synaptic coupling one can observe two successive 
dynamical transitions: one at small coupling from asynchronous to coherent dynamics and another
at larger inhibition from COs to asynchronous evolution. 
Furthermore, we show that the interval of synaptic couplings where COs are observable remains finite
in the thermodynamic limit.

The paper is organized as follows: Section II is devoted
to the introduction of the studied model and of the microscopic
and macroscopic indicators employed to characterize its dynamics.
The system is analyzed in Section III for a globally coupled topology,
where the WTA mechanism and the emergence of COs are discussed.
In Section IV, we study sparse random networks, with emphasis on
the role of current fluctuations to induce a rebirth in the neural activity at
large synaptic scale as well as their influence on collective behaviours.
The combined role of heterogeneity and delay on the dynamical behaviour of the system 
is addressed both for GCNs  (Sect. III)  and SNs (in Sect. IV). 
Section V deals with a detailed analysis of the  effect of disorder on finite
size networks. Finally, a brief discussion of the reported results can be found
in Section VI.

\section{Model -- Microscopic and Macroscopic Indicators}

We consider a heterogeneous inhibitory random network made of $N$ pulse-coupled leaky-integrate-and-fire (LIF) neurons. 
The evolution of the membrane potential of the $i^{th}$ neuron in the network, denoted by $v_{i}$, is given by:  
\begin{equation}
\label{model_membranepotential}
\dot{v_{i}}(t)=a_{i}-v_{i}(t)-\frac{g}{K}\sum_{n|t_{n}<t} S_{i,l(n)} \delta(t-t_{n}-t_{d})
\end{equation}
whenever $v_{i}$ reaches the firing threshold $v_{\theta}=1$ it is instantaneously reset to the resting value $v_{r}=0$ and a instantaneous $\delta$-spike is emitted at time $t_n$ and received by its post-synaptic neighbours after a delay $t_d$. 
The sum appearing in \eqref{model_membranepotential} runs over all the spikes received by the neuron $i$ up to the time $t$. 
$S_{i,l}$ denotes the connectivity matrix, with entries $1$, whenever  
a link connects the pre-synaptic neuron $l$ to the post-synaptic
neuron $i$, and $0$, otherwise. 
We consider both sparse (SNs) and globally coupled networks (GCNs). 
For sparse networks we randomly select the entries of 
$S_{i,l}$, however we impose that the number of pre-synaptic connections 
is constant and equal to $K$ for each neuron $i$, 
namely $\sum_{l \neq i}S_{i,l}=K$, since
autaptic connections are not allowed. Therefore, for the GCN we have $K=N-1$. The positive 
parameter $g$ appearing in \eqref{model_membranepotential} represents the 
coupling strength and the preceding negative sign denotes the inhibitory nature of the synapse. Each neuron is subject to a different supra-threshold input current $a_{i} > v_{\theta}$, representing the contribution both of the intrinsic neural excitability and of the external excitation due to projections of neurons situated outside the considered recurrent network. Heterogeneity in the excitabilities is introduced by randomly drawing $a_{i}$ from an uniform distribution of width
$\Delta a = a_2 - a_1$ defined over the interval $[a_{1}:a_{2}]$.
For simplicity, all variables are assumed to be dimensionless.

The microscopic dynamics can be characterized in terms 
of the interspike interval (ISI) $T_{i,ISI}$ statistics for each neuron $i$. 
The statistics is known once the corresponding probability density functon (PDF) $P(T_{i,ISI})$ 
is given, from which it can be obtained the average firing period $T_i = \langle T_{i,ISI} \rangle$ 
as well as the coefficient of variation $CV_i = \sigma(T_{i,ISI}) / \langle T_{i,ISI} \rangle$, being $\sigma(T_{i,ISI})$ the standard deviation of the $ISI$ distribution. The average firing rate of neuron $i$ is given by
$\nu_i = 1/T_i$. For the considered heterogeneous distribution of the
excitabilities, each isolated neuron is characterized by a different free spiking period, 
namely $T_{{\rm i,free}}=\ln [a_{i}/(a_{i}-1)]$. However, in the network the activity of each
neuron is modified by the the firing activity of its pre-synaptic neighbours.
In particular, the effective input $\mu_i$ to a generic neuron $i$ in the network can be written, 
within a mean-field approximation, as follows:
\begin{equation}
\label{effective_input}
\mu_i(t) = a_i - g [\nu(t)] n_A(t) \qquad ,
\end{equation}
where $[\nu(t)]$ is the average firing rate, with $[\cdot ]$  denoting the ensemble average over all the
neurons, and $n_A(t)$ is the percentage of active neurons.
A neuron will be supra- or below-threshold depending if $\mu_i(t)$ is larger or smaller than
$v_\theta$. The percentage of active neurons $n_A(t_f)$ in a certain time
interval $t_f$ is a quantity that we will employ to characterize the network at a
microscopic level. This is measured as  the percentage of neurons that have emitted at least two spikes within 
a time period $t_f$ after discarding a transient corresponding to the emission of $20 N$ spikes 
(we employ  these values for all the reported simulations, unless otherwise stated).
 
In order to study the collective behavior of the network we 
introduce an auxiliary field $E_i(t)$ for each neuron representing 
the linear superposition of the received train of spikes filtered opportunely.
In particular we filter each spike with a post-synaptic profile
having the shape of a $\alpha$-function $p(t)=\alpha^{2}t{\rm exp}(-\alpha t)$ ($t>0$), therefore the corresponding effective fields $E_{i}(t)$ 
can be obtained by integrating the following second order ordinary differential equations:
\begin{equation}
\label{model_fields}
\ddot E_{i}+2\alpha \dot E_{i}+\alpha^{2}E_{i}=\frac{\alpha^{2}}{K}\sum_{n|t_{n}<t}  S_{i,l(n)} \delta(t-t_{n}-t_{d}) \quad ;
\end{equation}
where $\alpha$ represents the inverse pulse width and it is fixed to $\alpha=$20.
The integration of the set of ordinary differential equations  \eqref{model_membranepotential} 
and \eqref{model_fields} has been performed in an exact manner by employing a refined event driven technique explained in details in \cite{politi2010lif}.

The macroscopic dynamics of the network can be analysed in terms of the
mean field 
$$
[E(t)] = \frac{1}{N}\sum_{i=1}^{N} E_{i}(t) \qquad ,
$$
which gives a measure of the instantaneous firing activity
at the network level. Furthermore, to identify collective oscillations 
it is more convenient to use the variance of the mean field $[E(t)]$ defined as
$$
\sigma^2([E])= \langle [E]^{2} \rangle- \langle [E] \rangle ^2
$$
where $\langle \cdot \rangle$ indicates the time average.

In general we will always measure either the time average or 
the variance of $[E]$, hence to avoid overuse of symbols and unless otherwise stated, $\langle E \rangle \equiv \langle [E] \rangle$ and $\sigma(E) \equiv \sigma([E])$.

\begin{figure}
\begin{center}
\includegraphics*[angle=0,width=0.95\linewidth]{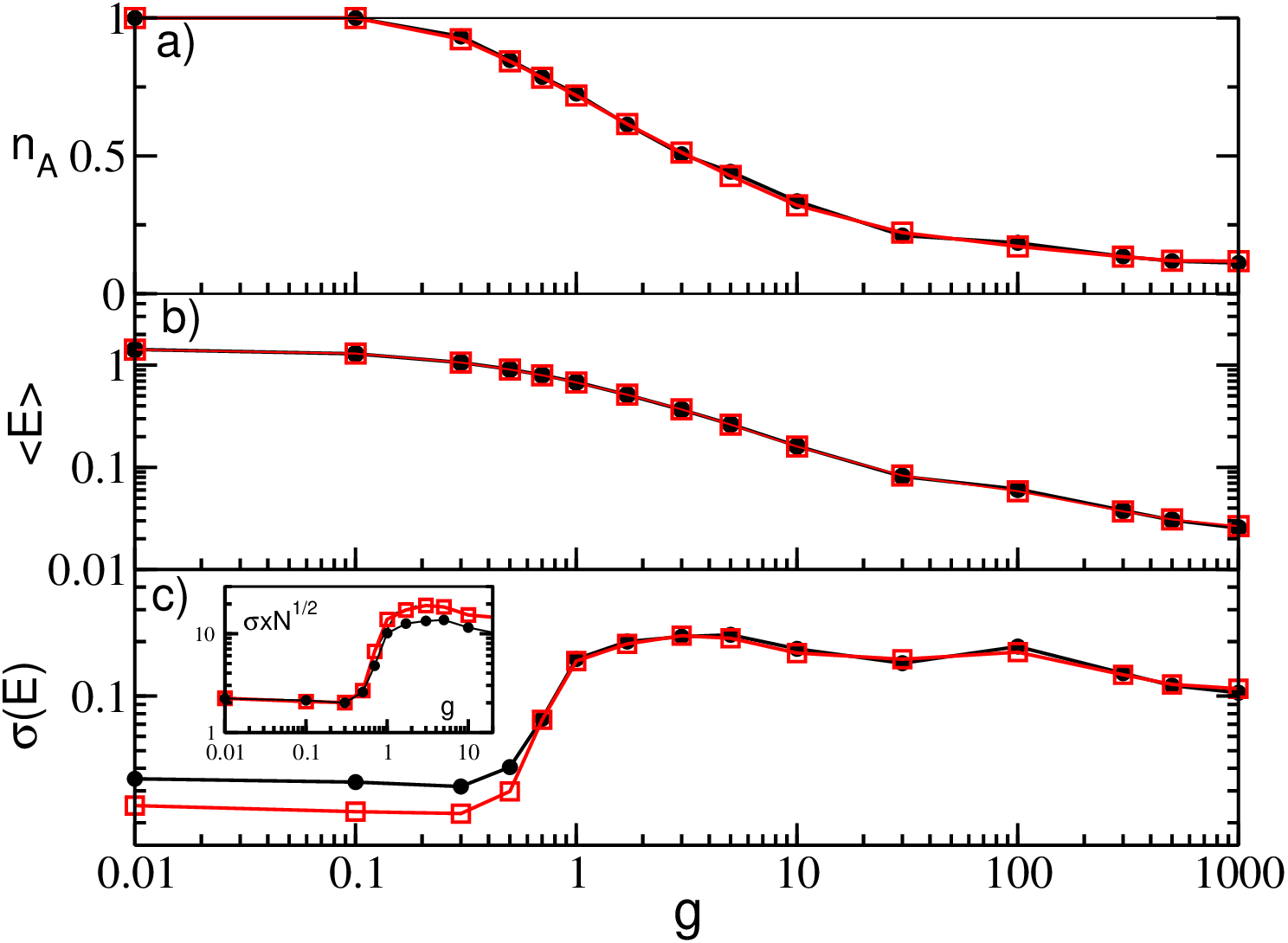}
\includegraphics*[angle=0,width=0.95\linewidth]{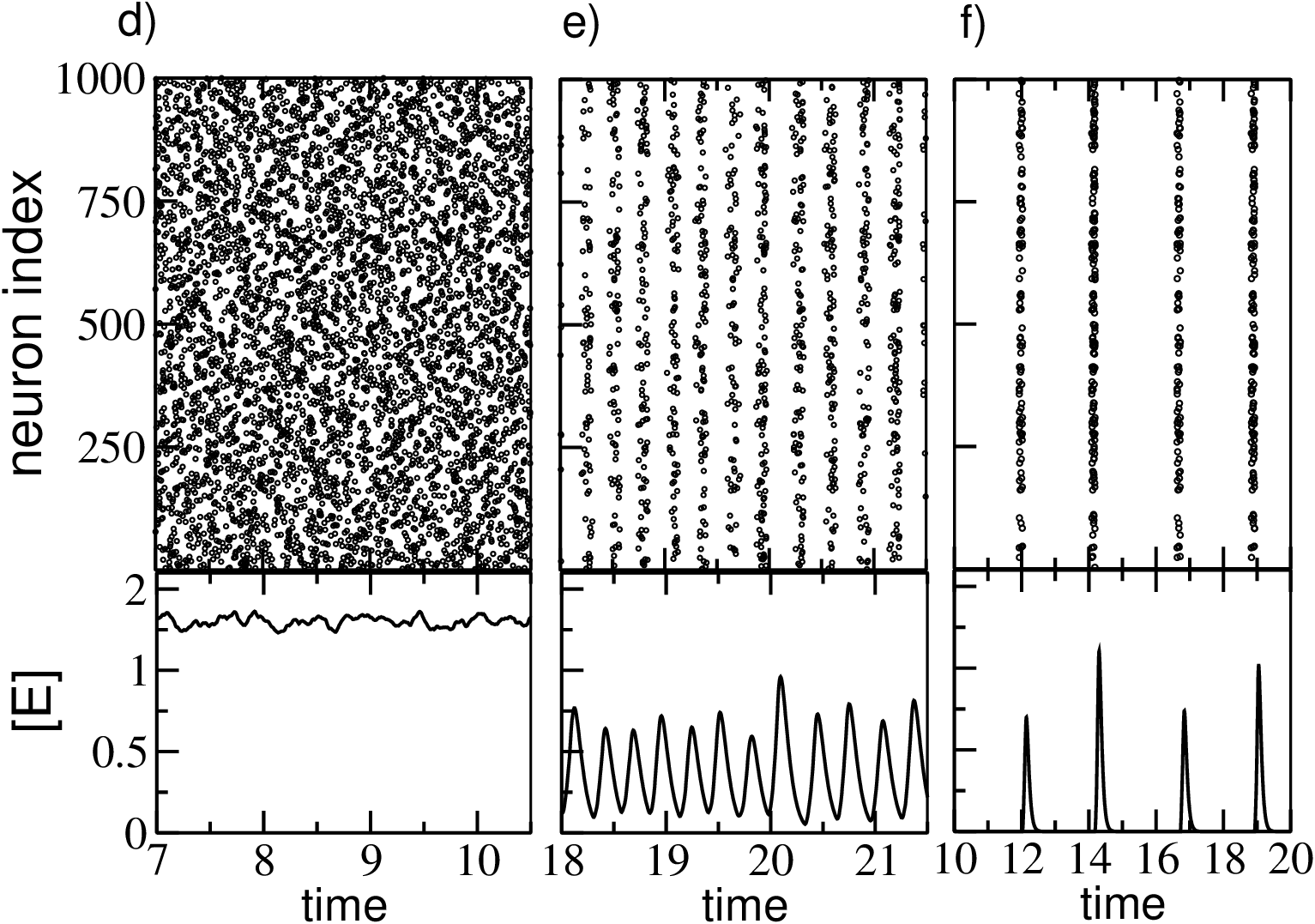}
\includegraphics*[angle=0,width=0.95\linewidth]{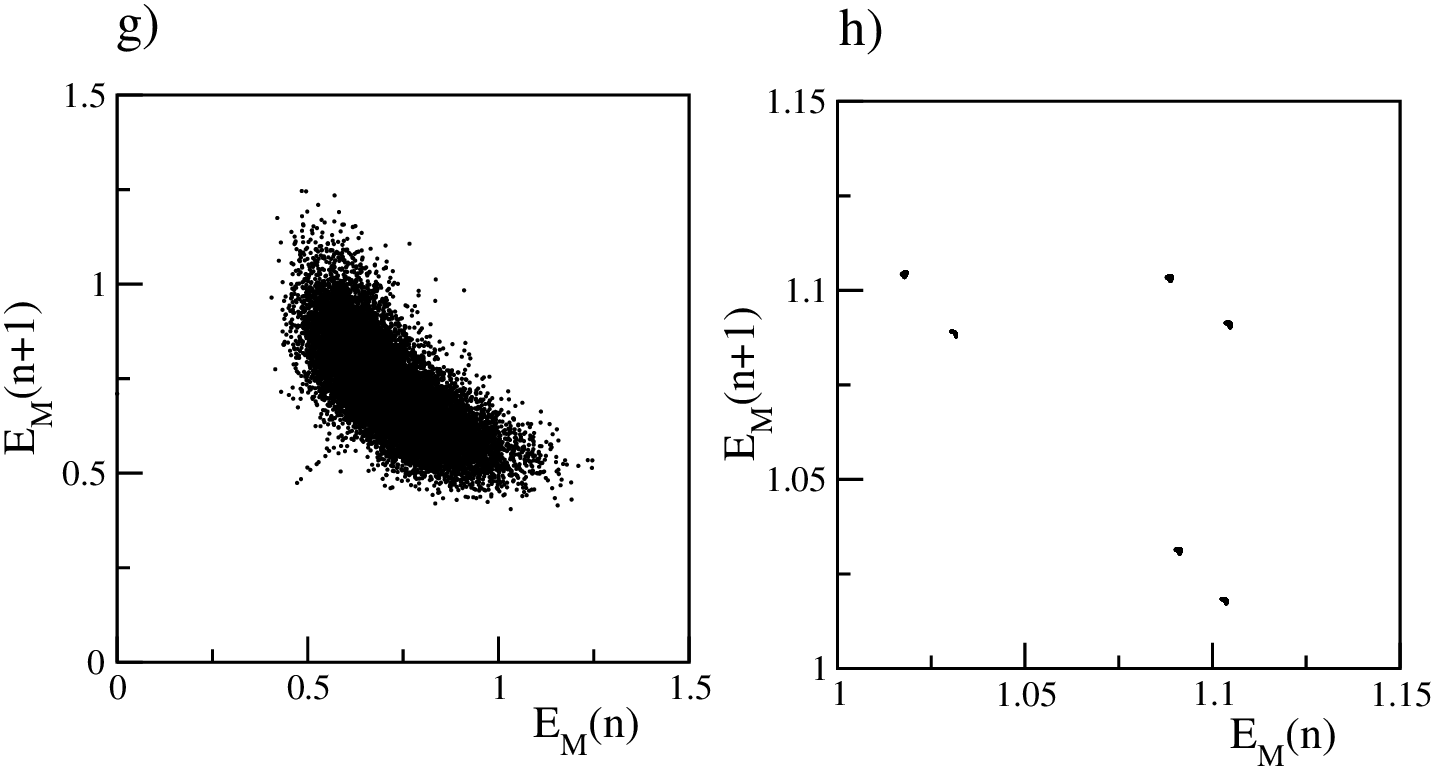}
\end{center}
\caption{\textbf{Winners take all in GCNs:} 
a) Fraction of active neurons $n_A$, 
b) time average of the field $\langle E \rangle$ and c) the corresponding fluctuations 
$\sigma(E)$ as a function of the strength of the inhibition $g$. 
In the inset in c) $\sigma(E)$ has been multiplied by $\sqrt N$. 
Black filled circles correspond to $N=4000$ while red empty squares represent $N=8000$. 
d-f) Raster plots (top) and  time traces of the field (bottom) for increasing 
coupling strength: from the left to the right $g=0.1$, $3$, and
$100$ for $N=4000$ (for reasons of clarity in the raster plots 
only the spikes of 1000 neurons are shown). 
g-h) Return maps for the maxima of the 
field $E_M$ for $g=3$ (g) and $g=100$ (h) in the case of $N=4000$.
Simulations in this figure were obtained after discarding a transient 
corresponding to $20 N$ spikes and  calculating the statistics over a 
time interval $t_f= 5 \times 10^2$ time units, apart for the data shown in panels g) and h) 
where $t_f= 5 \times 10^3$ time units. 
The data refer to a time delay $t_d = 0.1$, with $a_{1} = 1.2$ and $a_{2}=2.8$,
and $\alpha = 20$.} 
\label{fig:fig1}
\end{figure}

\section{Globally Coupled network}

First we will examine how the dynamics of a GCN will change for increasing synaptic coupling strengths $g$,
for a chosen time delay and a certain quenched distribution of the neuronal excitability.
The results of this analysis are reported in Fig. \ref{fig:fig1} (a-c) for two different system sizes, 
namely $N=4000$ and $N=8000$. Analogously to what found in absence of delay in \cite{angulo2017death_NJP}, 
we observe a steady decrease of the value of $n_A$ for increasing $g$ and  essentially no dependence
on the system size. Furthermore, the value of $n_A$ is independent of the value of the 
considered time period $t_f$ once a transient time is discarded.

For sufficiently small coupling all the neurons are active (i.e., $n_A =1$), and
the field $E$, which is a proxy of the firing activity of the network, presents an
almost constant value with few or none fluctuations. This indicates an
asynchronous activity \cite{olmi2012}, as confirmed by the raster plot shown
in Fig. \ref{fig:fig1} (d) for $g=0.1$. By increasing the coupling, $n_A$
reduces below one, because now the neuronal population splits in two groups :
one displaying a high activity, {\it the winners}, which are
able to mute the other group of neurons, {\it the losers},
which are usually characterized by lower values of the excitability $a_i$.
Further increases in the inhibition produces a steady decrease of 
the percentage of active neurons $n_A$ due to the increased
inhibitory action of the {\it winners} that induces an enlargement of 
the family of the {\it losers}  and an associated decrease in the 
network activity measured by $\langle E \rangle$ as shown in Fig.  \ref{fig:fig1} (b).
This is clearly an effect, which can be attributed to the WTA mechanism.

In \cite{ernst1995} it has been shown that
perfectly synchronized clusters of neurons emerge in homogeneous fully coupled inhibitory 
networks due to the transmission delay.
The presence of disorder (either in the excitability distribution or 
in the connections) leads to a smearing of the clusters associated to a non perfect 
synchronization~\cite{politi2010lif,Luccioli2010Irregular,Zillmer2009LongTrans} as we observe
in the present case. As shown in Fig. \ref{fig:fig1} (e) and (f),
for sufficiently large $g$, the emergence of the partially synchronized clusters produce collective 
oscillations at the macroscopic level. In particular, 
we observe a transition from an asynchronous state to collective oscillations,
as demonstrated in Fig.~\ref{fig:fig1} (c) by reporting $\sigma(E)$ as a
function of $g$ for different system sizes. At  $g \le 0.5$ $\sigma(E)$ tends to vanish as $1/\sqrt{N}$,
a typical signature of asynchronous dynamics, for larger values of the coupling strength 
(namely $g > 0.5$), $\sigma(E)$ displays a finite value independent of the system size signaling the presence 
of collective oscillations. 
The nature of these oscillations has been previously extensively 
analysed in~\cite{Luccioli2010Irregular}. In such study the authors 
have shown that at intermediate coupling strengths the collective dynamics is
irregular, despite the linear stability of the system, due to 
{\it stable chaos} mechanisms \cite{politi2010stable}.
This is evident from Fig.~\ref{fig:fig1} (e), where the first return
map for the maxima of the field $E_{M}(n)$ is reported
for $g=3$. In the present analysis we examine much larger
coupling strength than in \cite{politi2010stable}, namely
$g \gg 10$. At these large synaptic strenghts 
we observe that the complexity of the collectve dynamics reduces
due to the fact that the number of active neurons drastically 
declines, as shwon in Fig.~\ref{fig:fig1} (a).
The few active neurons have a quite limited spread in
their excitabilities (namely $\simeq n_A \Delta a$ \cite{note1})
thus promoting their reciprocal synchronization.
This is also evident from the sharp peaks in the mean field 
evolution (see Fig.\ref{fig:fig1} (f)) and from the periodic 
behaviour of the first return map of $E_M(n)$, shown in Fig.\ref{fig:fig1} (h).

\subsection{Role of the heterogeneity and the delay}

\begin{figure}
\begin{center}
\includegraphics[angle=0,width=0.95\linewidth]{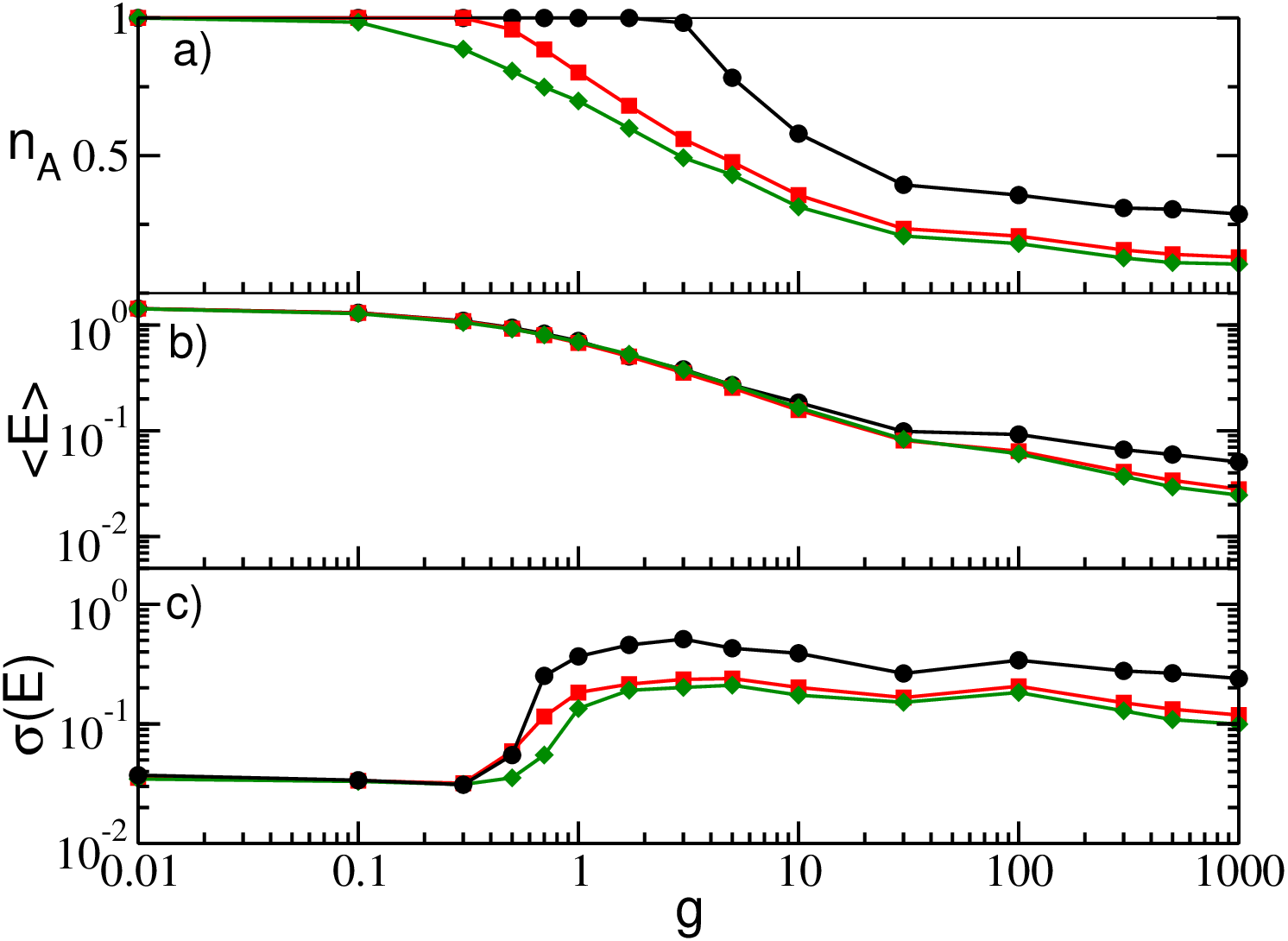}
\end{center}
\caption{\textbf{Relevance of the heterogeneity for the dynamics of GCNs:} 
From top to bottom, a) fraction of active neurons, b) average temporal 
value of the mean field and c) average fluctuations of the field $E$, 
for different strengths of inhibitory connections and several values 
of heterogeneity. Black circles refer to $\Delta a=$0.4, red squares to $\Delta a=$1.2 and green diamonds to $\Delta a=$1.8. A fixed time delay of $t_d = 0.1$ is used in this figure. The size of the network is $N=$4000, other parameters as in Fig.~\ref{fig:fig1}} 
\label{fig:fig2b}
\end{figure}

To better understand the influence on the dynamics of the parameters entering 
in the model, we considered different distributions of the excitabilities
and different time delays $t_d$. Let us first consider heterogeneity distributions
with different widths $\Delta a = a_2 - a_1$, but with the same average
value $[a]=2$, for a fixed value of the delay (namely, $t_d = 0.1$).
As shown in Fig. \ref{fig:fig2b} (a) and as already 
demonstrated in case of absence of delay \cite{angulo2017death_NJP} for 
$a_1 \to v_\theta = 1$ (corresponding to $\Delta a \to 1$ in the present case)
any arbitrary small amount of inhibition is sufficient to induce
neuronal deactivation. However, for increasing values of $a_1$
(for decreasing widths $\Delta a$)
the onset of neuronal deactivation occurs at increasingly
larger $g$-values, since larger amount of inhibition
are required to silence the neurons with smallest excitability.
This also explains why the values of the curves $n_A=n_A(g)$ 
decrease for increasing $\Delta a$, as shown in Fig. \ref{fig:fig2b} (a).
 
The average mean field does not present significant modifications with $g$ as seen in 
Fig.~\ref{fig:fig2b} (b). This is due to the fact that, from a mean field perspective, 
the system is subject to the same average excitability and hence one does not expect large 
deviations in the average firing rates. There are only small 
deviations at very large $g$ where the network with wider dispersion in the
excitabilities display slightly smaller firing rates, just because the number
of active neurons is drastically decreased. 
This effect is much more evident in Fig.~\ref{fig:fig2b} (c), 
where we can observe that the value of $\Delta a$ significatively affects both the onset and 
the amplitude of the collective oscillations as measured by $\sigma(E)$.
This because the decrease of $n_{A}$ brings to a 
reduction in the number of partially synchronized 
neurons and in turn of the amplitude of the fluctuations of the field.

\begin{figure}
\begin{center}
\includegraphics[angle=0,width=0.95\linewidth]{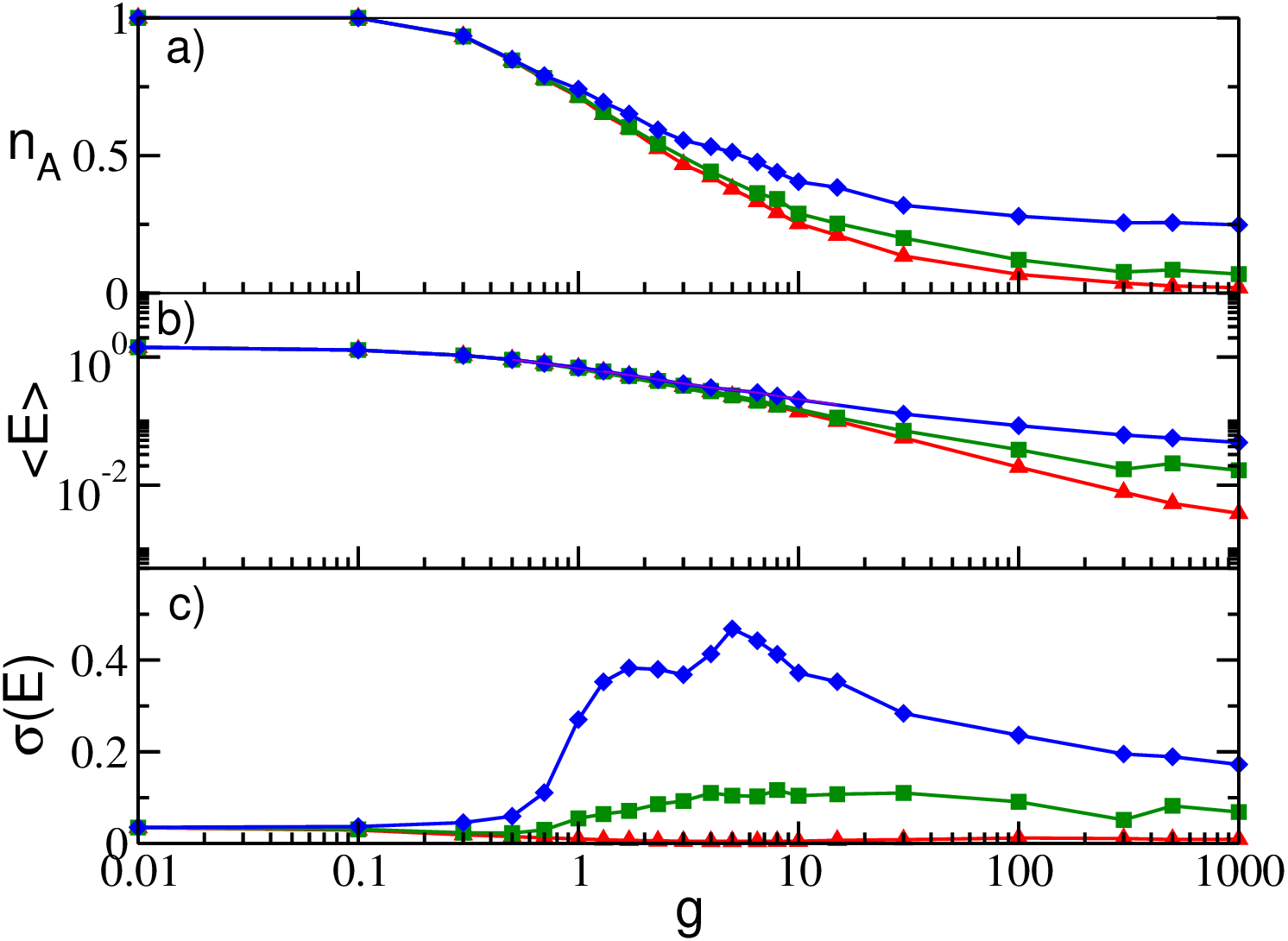}
\end{center}
\caption{\textbf{Relevance of the time delay in GCNs:} 
From top to bottom a) Fraction of active neurons, b) Average temporal 
value of the mean field and c) average fluctuations of the field $E$, 
for different strengths of inhibitory connections and
several values of delay, $t_d=$0.25 (blue diamonds), $t_d=$0.05 (green squares) 
and $t_d=$0.005 (red triangles). 
For this figure a fixed heterogeneity distribution with $a_{1} = 1.2$ and $a_{2}=2.8$ is used. 
The size of the network is $N=$4000, other parameters as in Fig. \ref{fig:fig1}} 
\label{fig:fig2a}
\end{figure}

Let us now consider the influence of the time delay, for a fixed heterogeneity distribution. From Fig.~\ref{fig:fig2a} (a), it appears that $n_A$ approaches 
an asymptotic plateau for very large coupling, whose value steadily decreases
for increasing $t_d$. Indeed, for very small delays the survivors 
reduce to few units, e.g. see the example reported in Fig.~\ref{fig:fig2a} (a)
for $t_d = 0.005$. This dependence of $n_A$ on the delay at large couplings is
confirmed also by the corresponding values of the field $E$ shown in 
Fig.~\ref{fig:fig2a} (b). Overall, longer synaptic delays counteract the effect of the heterogeneity and therefore the neural deactivation, as a matter of fact for $t_{d} \simeq T_{{\rm free}} \simeq 1$ the fraction of 
active neuron is $\simeq 0.67$ still for g=1000 (data not shown).

Furthermore, the delay has a crucial role in the emergence
of collective oscillations that can be observed already
for extremely small delay, e.g $t_d \geq 0.05$ for the parameters considered in
Fig.~\ref{fig:fig2a} (c). Indeed, no collective
oscillations have been observed in heterogeneous GCN in absence of delay
at any coupling stregth \cite{angulo2017death_NJP}.
By increasing the delay we observe larger and larger 
oscillations in the field $E$, as shown by the curves reported in
Fig.~\ref{fig:fig2a} (c). 

These behaviours can be explained by the fact that collective oscillations
are due to the presence of clusters of neurons at a microscopic level.
As shown in \cite{ernst1995,ernst1998} for homogeneous 
systems and confirmed in \cite{politi2010lif} for heterogenous
networks the average number of clusters $N_c$ increases proportionally
to the inverse of the delay. Therefore for small delay we
expect to observe an asynchronous state, characterized by $N_c \simeq N$,
while for increasing delay $N_c$ decreases and thus the neurons are
more and more synchronized, thus promoting larger collective fluctuations.
The increase in the overall synchronization leads to a reduced effective
variability in the neuron dynamics, thus preventing neuronal deactivation.
Indeed, disorder promotes deactivation as demonstrated in \cite{angulo2017death_NJP} in absence of delay and as shown in Fig.~\ref{fig:fig2b} (a), where $n_A$ is reported for various $\Delta a$ values.

\section{Sparse Network}

\begin{figure}
\begin{center}
\includegraphics[angle=0,width=0.95\linewidth]{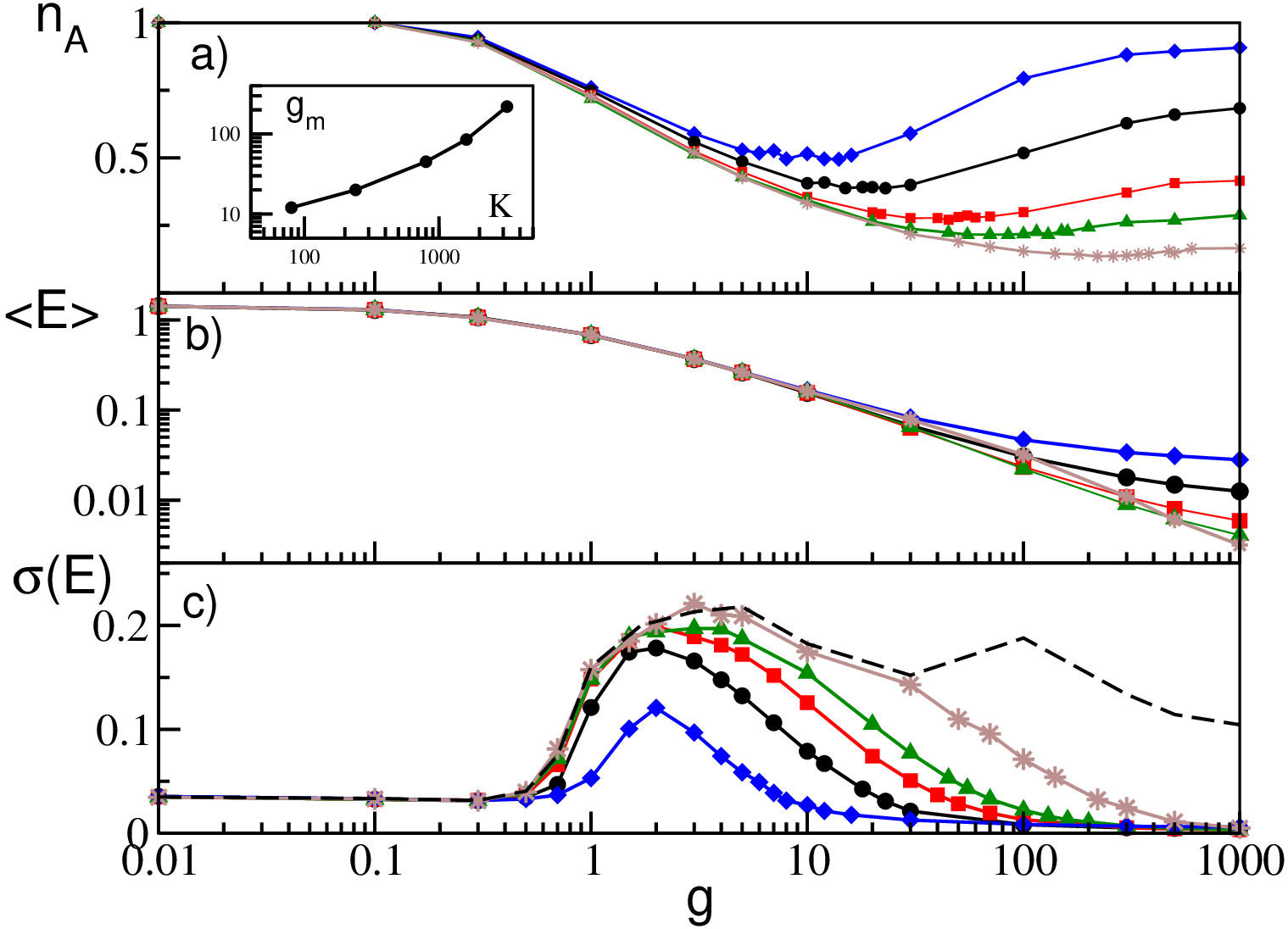}\\
\vspace{5mm}
\includegraphics[width=0.95\linewidth]{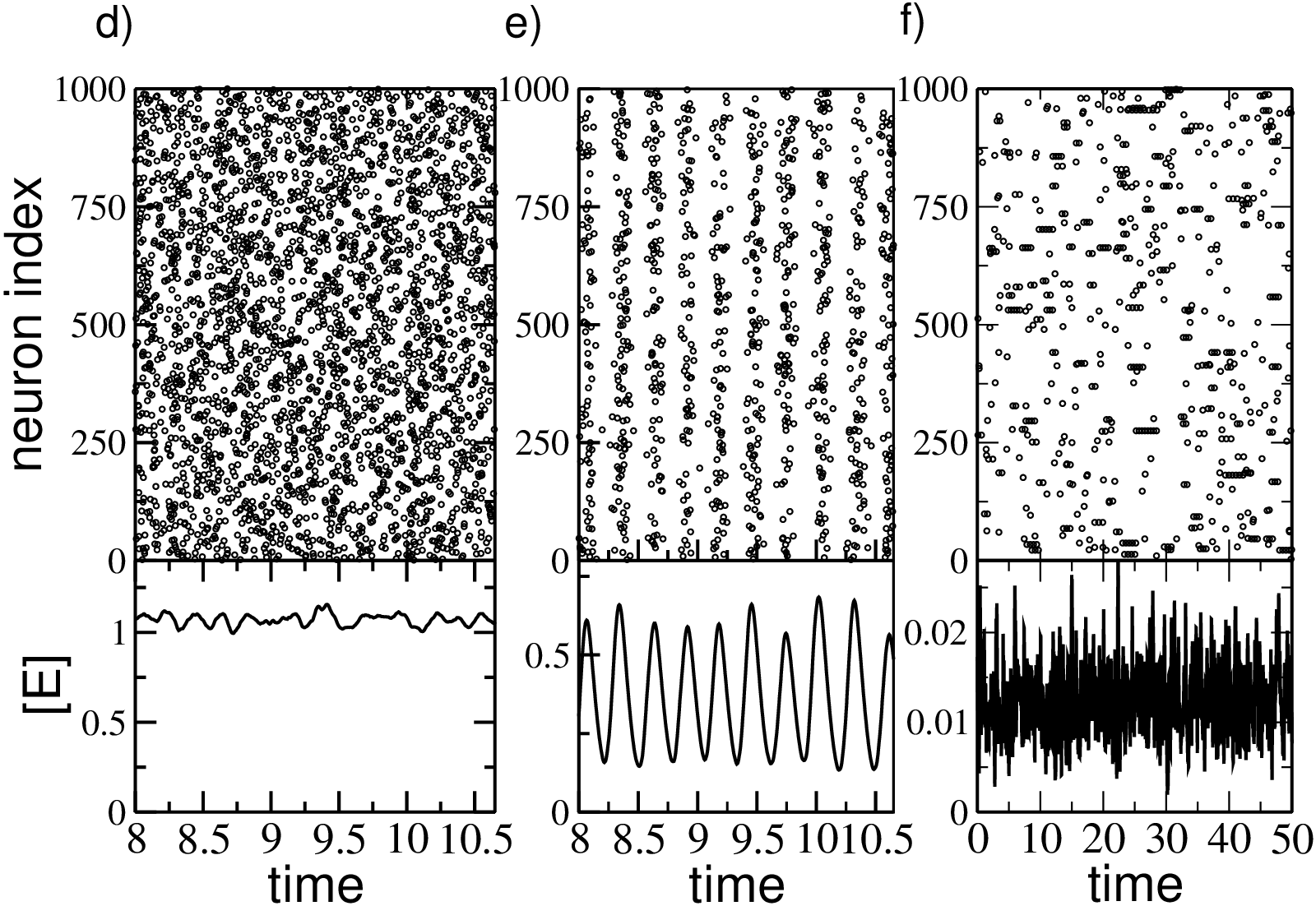}
\end{center}
\caption{\textbf{Reactivation and collective oscillation in SNs:} 
a) Fraction of active neurons $n_A$, b) average mean
field $\langle E \rangle$ and c) fluctuations $\sigma(E)$ of the mean field as a function 
of the inhibition. The data refer to a fixed network size $N=$4000 and different in-degrees:
namely, $K=$80 (blue diamonds), $K=$240 (black circles), $K=$800 (red squares), $K=$1600 
(green triangles) and  $K=$3200 (brown stars). In the inset of a) the value of $g_{m}$ is plotted 
versus the in-degree $K$, where $g_{m}$ has been estimated by using a cubic regression in the 
region of the minimum of $g$. The black dashed curve in c) refers to the 
GCN previously reported in Fig.~\ref{fig:fig1} (c).
d-f) Raster plot (top) and time course of the mean field $[E]$ (bottom)
for $g=0.3$ (d), $g=3$ (e) and $g = 1000$ (f). Other parameters as in Fig. \ref{fig:fig1}} 
\label{fig:fig3}
\end{figure}

We will now consider the diluted case, i.e. each neuron has now
exactly $K < (N-1)$ random pre-synaptic neighbours. In this
case we observe that $n_A$ has a non-monotonic behaviour
with $g$, as shown in Fig. \ref{fig:fig3} (a) for
different values of the in-degree $K$ for a fixed system size,
namely $N=4000$. In particular, we observe that for small coupling
$n_A$ shows a decrease analogous to the one reported for the GCN,
hower for for synaptic coupling larger than a critical value $g_m$ 
the percentage of active neurons increases with $g$. 
This behaviour indicates that large inhibtory coupling
can lead to a reactivation of previously inactive neurons
in sparse networks, as prevously reported 
in inhibtory networks in absence of delay for conductance based 
\cite{ponzi2013optimal} and LIF neuronal models \cite{angulo2015Striatum, angulo2017death_NJP}.
The value of $g_m$ grows faster than a power-law
with the in-degree $K$, as shown in the inset of Fig.~\ref{fig:fig3} (a). 
In particular, we expect that for $K \to (N-1)$,
i.e. by recovering the fully coupled case,
$g_m \to \infty$ and $n_A$ converges towards
the curve reported in Fig.~\ref{fig:fig1} (a).

Analogously to GCNs the average mean field $\langle E \rangle$
is steadily decreasing with $g$ indicating that 
the neuronal dynamics slow down for increasing inhibition
and that for sufficiently
large $g$ all the neurons can eventually be reactivated but with
a definitely low firing rate. The dependence of $\langle E \rangle$
 on $K$, shown in Fig. \ref{fig:fig3} (b), reveals that for
$g < g_m$ essentially all curves coincide, while at larger synaptic
coupling the smaller is $K$ the larger is the value of the field,
this behaviour is clearly dictated by that of $n_A$. More neurons 
are reactivated, higher is the filtered firing rate measured by $\langle E \rangle$.

The mean field fluctuations $\sigma(E)$ have now a striking different behaviour
with respect to the GCN, because now $\sigma(E)$ displays a maximum at some
intermediate $g$ value, while for small and large coupling $\sigma(E)$ tends to vanish, 
as shown in Fig.~\ref{fig:fig3} (c). 
This suggests that collective oscillations 
are present only at intermediate coupling, while out of this range the dynamics is asynchronous. 
This is confirmed by the raster plots and the fields
reported in Fig.~\ref{fig:fig3} (d-f) for different synaptic couplings.
At small coupling (namely, $g=0.3$) a clear asynchronous state, characterized
by an almost constant $[E]$, is observable , while in an intermediate 
range of synaptic couplings clear collective oscillations are present, as testified by the raster plot 
and the field shown in Fig.~\ref{fig:fig3} (e) for $g=3$.  
Furthermore, for sufficiently large coupling the asynchronous dynamics is characterized by 
a sporadic bursting activity with an intra-burst period corresponding to 
to $T_{{free}}$ of the considered neuron, as shown for $g=1000$ in Fig.~\ref{fig:fig3} (f).

From Fig. \ref{fig:fig3} (c) it is also evident
that the onset and the amplitude of the collective dynamics strongly depend 
on the dilution measured in terms of the in-degree $K$, and that 
for $K\rightarrow (N-1)$ the globally coupled behavior is recovered. 
In particular, smaller is $K$ smaller is the amplitude of the collective oscillations and 
narrower is the synaptic coupling interval where they are observable. 
These two effects are due to the fact that the disorder in the connectivity
distribution increases as $1/\sqrt{K}$. Therefore, on one side the clusters
of partially synchronized neurons, which are responsible for the 
collectivve oscillations, are more smeared at smaller $K$ 
thus inducing smaller amplitudes of the oscillations. 
On the other side, the disorder prevent the emergence of
collective oscillations, thus the region of existence is reduced at lower $K$.
As a matter of fact, for $N=$4000 we start to observe collective 
oscillations only when $K>40$. The existence of a critical connectivity for the 
emergence of collective dynamics is a general feature of sparse
networks \cite{Luccioli2012PRL,di2018}.

\begin{figure}
\includegraphics[width=0.95\linewidth]{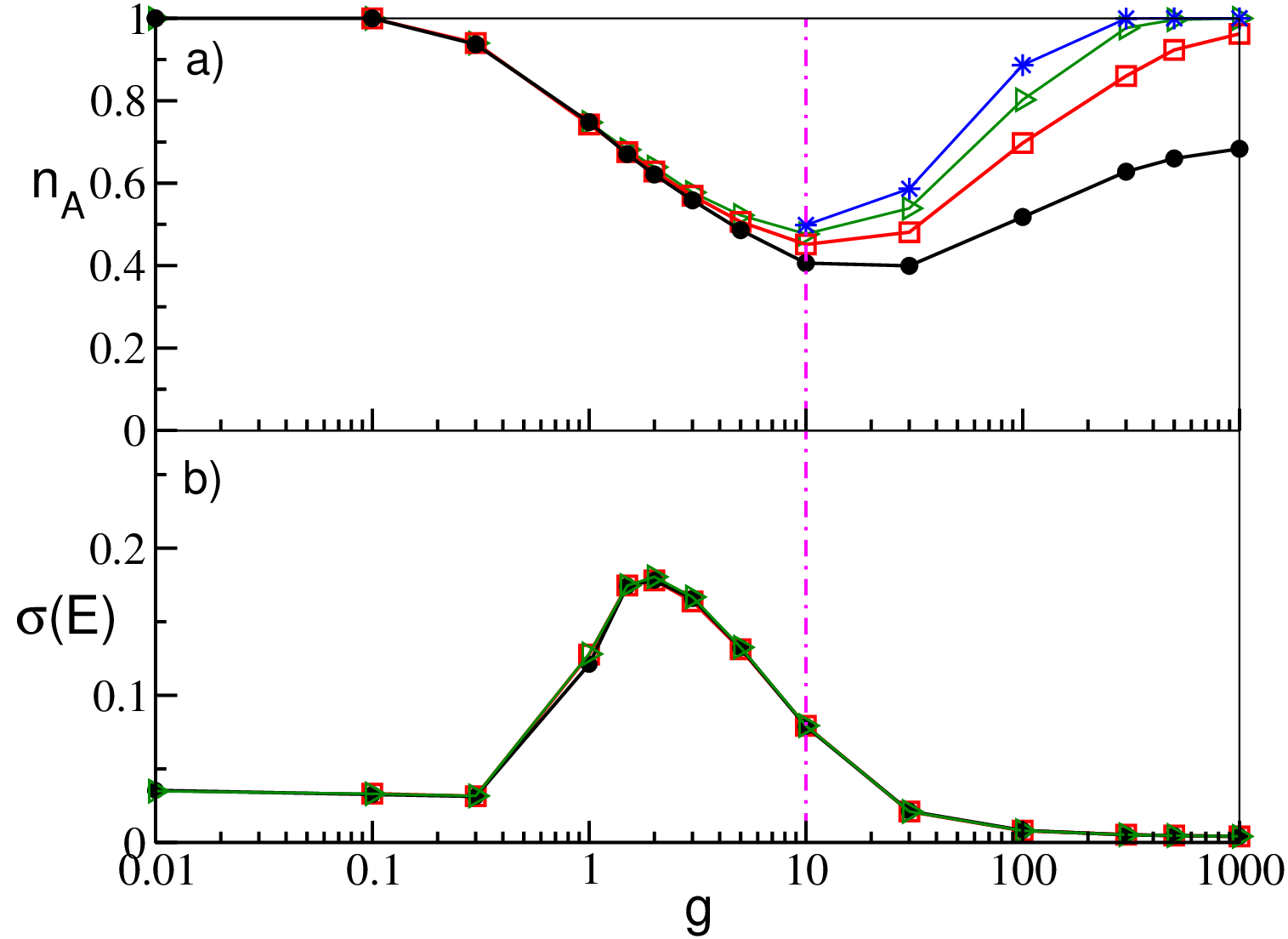}
\caption{\textbf{Finite time effects in SNs:} 
a) Fraction of active neurons as a function of inhibition for four different time windows,
namely $t_f = 5\times 10^2$ (black circles), $t_f=5\times 10^3$ (red squares),  $t_f=5 \times 10^4$ (green triangles)
and $t_f=5 \times 10^5$ (blue stars). b) Fluctuations of the mean field as a function of inhibition 
estimated for the three shortest time windows $t_f$ reported in panel a). 
The vertical dot-dashed magenta line denotes $g_m$, as estimated for $t_f=5 \times 10^5$.
The data refer to $N=4000$ and $K = 240$, all the other parameters as in Fig.~ \ref{fig:fig1}} 
\label{fig:Time_Dependence}
\end{figure}

In order to understand if the value $t_f$ of the time interval over which we measure $n_A$ and $\sigma(E)$
has an influence on the observed effects, let us examine the dependence of
these two quantities on $t_f$ for a fixed size $N$ and in-degree $K$.
The results of this analysis reported in Fig. \ref{fig:Time_Dependence} 
show that for $g < g_{m}$ the percentage of active neurons
is almost insensible to the considered time window, similarly
to what observed for the GCN. On the other hand, for $g > g_{m}$
the value $n_A$ grows with $t_f$ and for sufficiently long times
and for sufficiently large $g$ eventually all the neurons can 
be reactivated. However, as shown in Fig. \ref{fig:Time_Dependence} (a) 
the growth $n_A$ noticeably slow down for increasing $t_f$
and we can safely affirm that for $t_f > 10^5$ the further evolution of $n_A$
occurs on unrealistic long time scales.   
For what concerns $\sigma(E)$ finite time effects are essentially
not present, as shown in Fig. \ref{fig:Time_Dependence} (b).

\subsection{The role of current fluctuations}

Previous analysis of inhibitory networks in absence of delay \cite{ponzi2013optimal,angulo2015Striatum,angulo2017death_NJP} 
have clearly shown that the position $g_m$ of the minimum of $n_A$ marks the
transition from a regime dominated by the activity of the supra-threshold 
neurons ({\it mean driven}) to a regime where the most part of the neurons are below threshold
and the firing is mainly due to current fluctuations ({\it fluctuation driven}) \cite{renart2007}.

\begin{figure}
\includegraphics[width=0.95\linewidth]{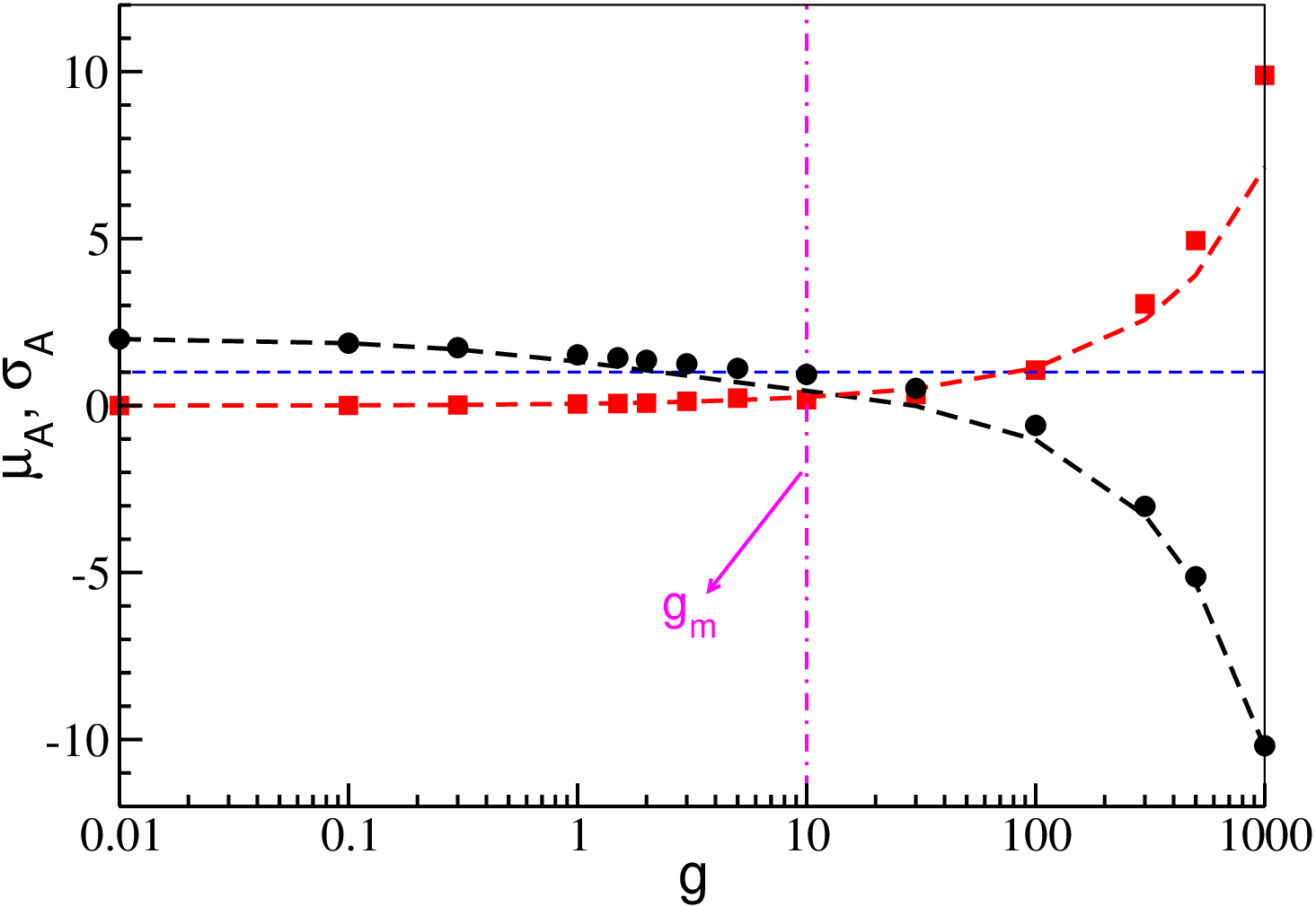}
\caption{\textbf{Mechanism of neural reactivation in SNs:} Average $\mu_A$ (black) and 
fluctuations $\sigma_A$ (red) of 
the effective input current in the active sub-population.  The dashed curves are 
the theoretical estimations (Eqs.~\eqref{eq:MuSigma} and ~\eqref{eq:Sigma}) while the filled symbols 
represent the numerical estimations. 
The firing threshold of the LIF neuron 
is depicted as a blue dashed line. Vertical dot-dashed line line denotes the measured value $g_m$.
A SN of size $N = 4000$ and $K = 240$ is considered; all other parameters as in Fig.~\ref{fig:fig1}.} 
\label{fig:Neuron_reactivation}
\end{figure}

In order to verify if also in the present case the origin of the neuronal reactivation
is related to such a transition, let us analyze the system at a mean-field level,
in this framework the activty of a neuron is completely determined
by the average input current and by its fluctuations. 
Let us limit our analysis to the active neurons, since these are the only
ones contributing to the network dynamcs, in particular 
the average effective input to the active neurons can be estimated as follows
\begin{equation}
\label{eq:MuSigma}
\mu_A  =  I_{A} - g  \nu_{A}n_{A} 
\end{equation}
where $I_A$ ($\nu_A$) is the average excitability (firing rate) of the active neurons. 
The current fluctuations can be estimated by following \cite{tuckwell2005, BrunelHakim1999},
despite the dynamics is fully deterministic, 
thanks to the disorder induced by the sparsness in the connections each neuron can be seen 
as subject to $n_A K$ uncorrelated Poissonian
trains of inhibitory spikes of constant amplitude $g$ and characterized by an average 
firing rate ${\nu_{A}}$ . Therefore the current
fluctuations can be estimated as follows: 
\begin{equation}
\sigma_A = g\sqrt{\frac{ \nu_{A} n_{A}}{K}} \qquad .
\label{eq:Sigma}
\end{equation}
As it can be appreciated from Fig.~\ref{fig:Neuron_reactivation},
the theoretical estimations ~\eqref{eq:MuSigma} and ~\eqref{eq:Sigma} (dashed curves)
are in very good agreement with the numerical
data for $\mu_A$ and $\sigma_A$ (filled symbols) over the whole considered range of the 
synaptic inhibition (corresponding to five orders of magnitude).
In the same figure, one observes a  steady decrease of $\mu_A$ with $g$, which can be understood 
from its expression \eqref{eq:MuSigma}, since $\nu_{A}$ is a quantity monotonically decreasing with 
the synaptic strenght despite the neural reactivation present in SNs.
This can be inferred from the behaviour of the mean field $[E]$, which
is strictly connected to $\nu_{A}$, reported in Fig.~\ref{fig:fig3} (b).
 On the other hand, the fluctuations of the input currents increase with $g$, thus indicating that
in \eqref{eq:Sigma} the growth of $g$ prevails over the decrease of $\nu_A$.

The key result explaining the mechanism behind neural reactivation is reported
in Fig. \ref{fig:Neuron_reactivation}:  it is clear that $\mu_A$ becomes smaller than the firing threshold $v_{\theta}=1$
exactly at $g = g_m$, in concomitance with a dramatic growth of the current fluctuations. Therefore for $g > g_m$, since all the neurons are on average below threshold, the neural firing is mostly due to current fluctuations
and not to the intrinsic excitability of each neuron. 
Therefore we expect that for large coupling strength, on one side the average firing of the neurons 
will become slower, as indeed shown in Fig. \ref{fig:fig3} (b), while on the other side
the fraction $n_A$ of firing neurons will increase,  thanks to the increase of $\sigma_A$ with $g$.
Therefore we can confirm that the occurrence of the minimum in $n_A$ signals 
a transition from a mean driven to a fluctuation driven dynamics, analogously 
to what found in \cite{angulo2017death_NJP} in absence of delay.
 
However, in the present case current fluctuations have also a destructive role on the collective dynamics
induced by the delay. As it can be inferred from Fig.~\ref{fig:Time_Dependence} (b)
(see also Fig.~\ref{fig:FiniteSize} in the last section), the coherent motion
disappears as soon as $g > g_m$ due to the random fluctuations in the input currents
which completely desynchronize the neural activity.

\subsection{Characterization of the microscopic dynamics}

The analysis of the microscopic dynamics can clarify the different
observed regimes. In particular, we will consider the
dynamics of the neurons in a network with $N=4000$ and $K=240$ in three 
typical regimes: namely, in presence of collective motion ($0.5 <g < g_m$),
in proximity of the minimum of $n_{A}$ ($g \simeq g_m = 10$) and 
for very large inhibition ($g >> g_m$).
In particular, for each synaptic coupling we study the distribution of the ISIs, $P(T_{ISI})$,
for three representative neurons characterized by high 
(H), intermediate (I) and low (L) average firing rates.

\begin{figure*}
\includegraphics[angle=0,width=0.4\linewidth]{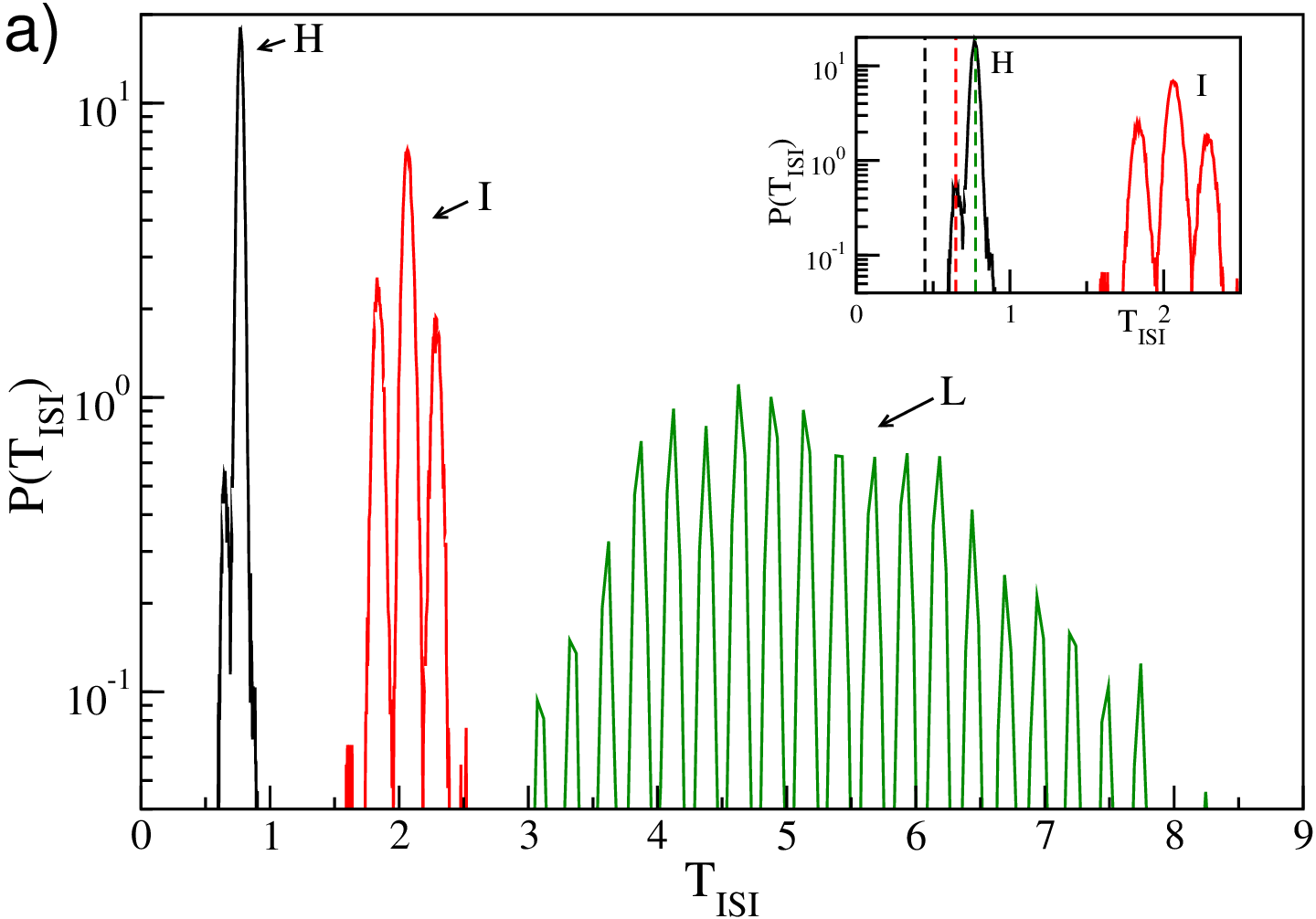}
\includegraphics[angle=0,width=0.4\linewidth]{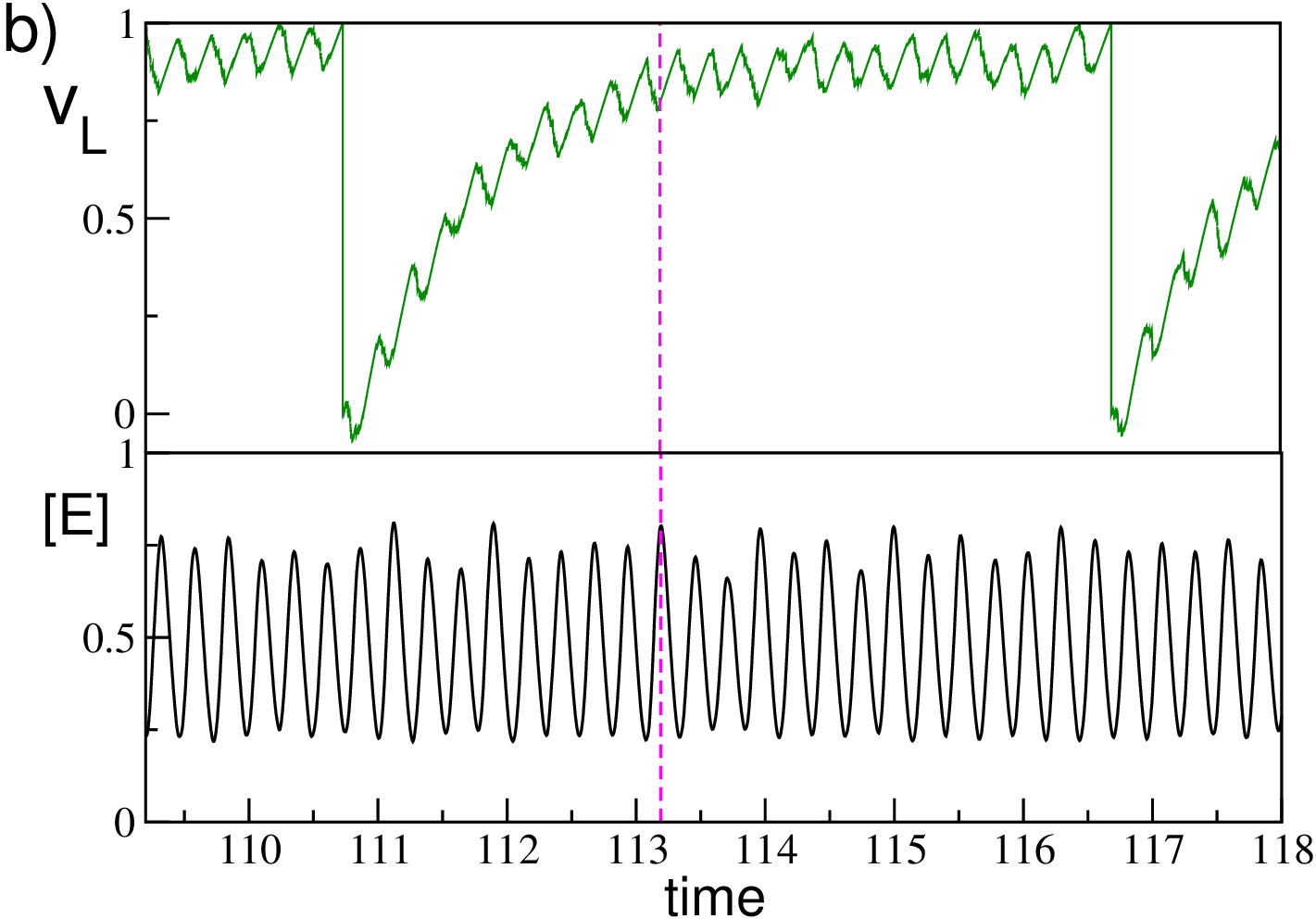}
\includegraphics[angle=0,width=0.4\linewidth]{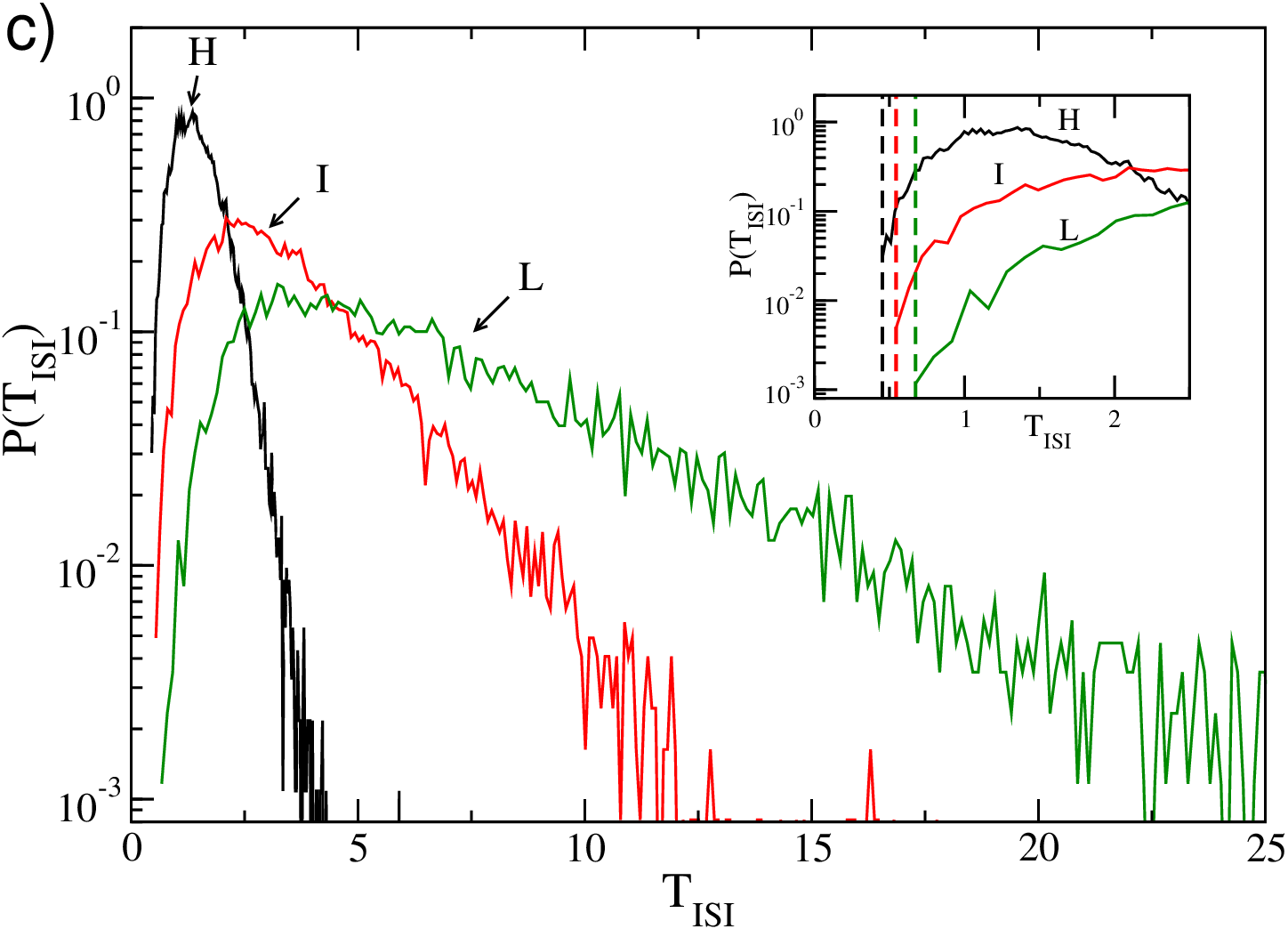}
\includegraphics[angle=0,width=0.4\linewidth]{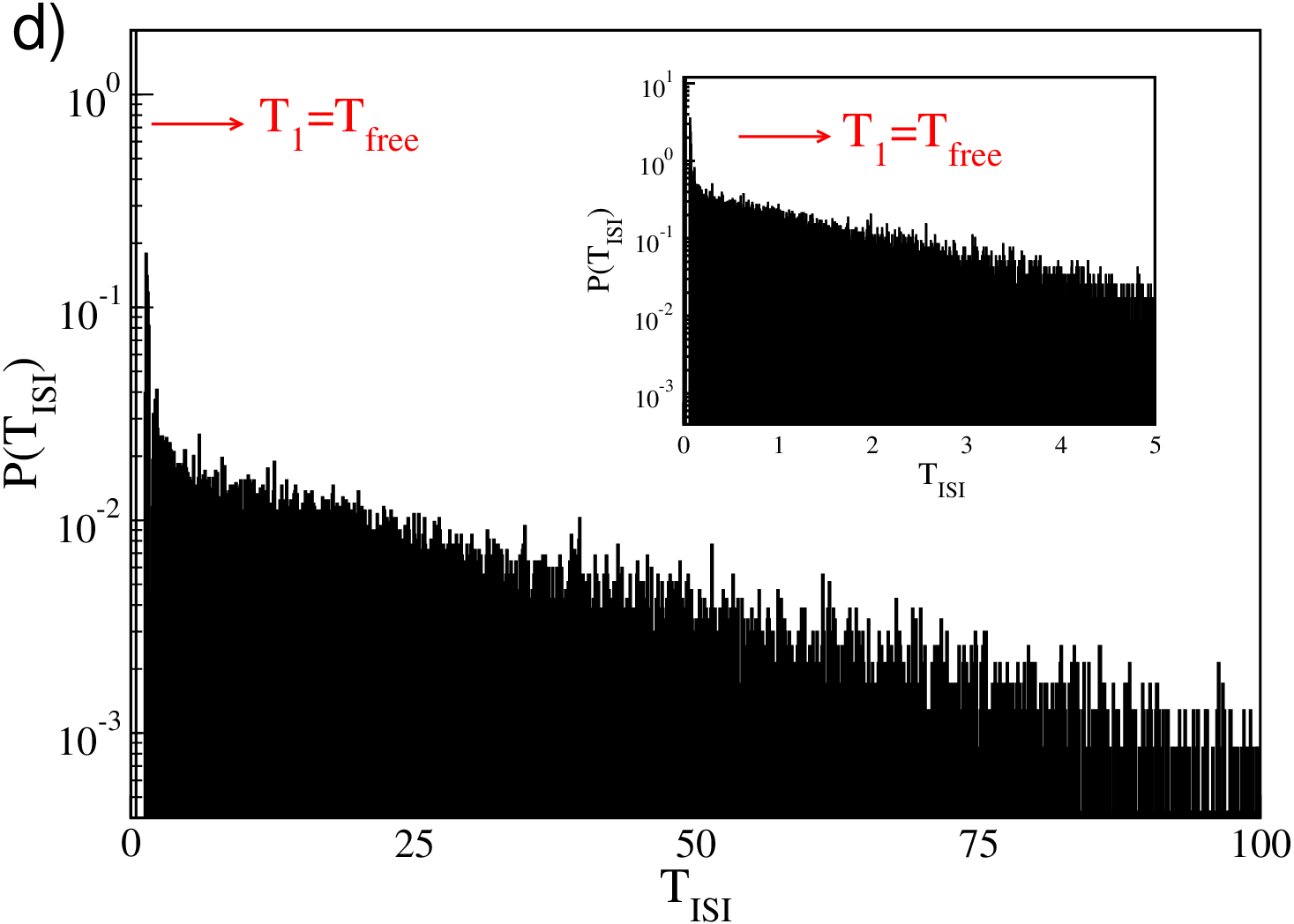}
\caption{{\bf Microscopic behavior: ISI analysis in SNs.} Probability 
distributions of the ISIs, $P(T_{{\rm ISI}})$, of representative neurons 
for increasing values of the coupling strength:
namely, $g=2$ (a), $g=30$ (c) and $g=1000$ (d). In (a) and (c) the 
black, red and green curves correspond to neurons with high (H), intermediate (I) 
and low (L) average firing rate; in the inset it is shown a close-up where 
the free periods of the three neurons, $T_{{\rm free}}$, are indicated 
by vertical dashed lines with the same color code. Note that starting from the coupling strength $g=30$ the 
value $T_{{\rm free}}$ appears as the first channel of the histograms. In (d) the first peak of the 
histogram, $T_{{\rm 1}}$, corresponds to the period of the free neuron $T_{{\rm free}}$. 
In (b) it is shown the 
mechanism originating the multi-peak structure in the distributions of panel (a):   
it is represented on the same time axis an instance of the membrane potential 
$v_{L}$ of the neuron with low firing rate and of the field $[E]$ (the vertical dashed line 
marks the locking between local minima of $v_{L}$ and local maxima of $[E]$). 
Network size $N=4000$ and $K=240$. Other parameters as in Fig. 1.} 
\label{fig:Bursting}
\end{figure*}

The results of this analysis are reported in Fig.~\ref{fig:Bursting}, where 
we considered $g=2$ (a,b), $g=30$ (c) and $g=1000$ (d).
In particular, $g=2$ corresponds to the maximum in the 
amplitude of the collective oscillations measured by $\sigma(E)$ 
(see Fig.~\ref{fig:Time_Dependence}). For this synaptic coupling
the distrbutions $P(T_{\rm ISI})$ are quite peculiar, being characterized
by several peaks separated by a constant time lag $\delta t$. 
The number of peaks and the value of $\delta t$ increase going from 
the fastest to the slowest neuron:
namely, the time lag $\delta t$ varies from $\sim 0.12$ (F)
to $\sim 0.23$ (I) and $\sim 0.25$ (L). 

This structure can be traced back to the coherent inhibitory action of clusters of partially 
synchronized neurons, coarse grained by the collective field $[E]$, on the targeted neuron. 
This effect is shown in Fig.~\ref{fig:Bursting} (b), where on the same
time interval are reported the membrane potential $v_{L}$ of the neuron with low firing rate 
and $[E]$. The average field $[E]$ displays irregular oscillations due to 
the clustered activity of the neurons, furthermore it is clear that the occurrence of every local maximum 
in $[E]$ is in perfect correspondence with a local minimum of  $v_{L}$. 
Therefore, a spike can be emitted only in correspondence of a local minimum of the field,
thus the $T_{{\rm ISI}}$ of neuron (L) should be multiples of the oscillation period of 
$[E]$, which is $\sim 0.25$. 
The locking with the collective field is progressively less effective for the neurons 
with higher firing rates (namely, (I) and (H)) and this reduces the multi-peak 
structure and the value of $\delta t$. 
Moreover, in this regime dominated by collective inhibitory oscillations the
minimal $T_{ISI}$ for each neuron is definitely larger than the corresponding $T_{\rm free}$.
Obviously, the more active the neuron is, 
the closer to $T_{{\rm free}}$ is the minimal $T_{ISI}$ (see the inset in Fig.~\ref{fig:Bursting} (a)).

For larger values of $g \ge g_m$, the collective oscillations vanish 
and accordingly the multi-peak structure in $P(T_{\rm ISI})$ disappears,
while the statistics of the firing times
becomes essentially Poissonian as shown in Fig.~\ref{fig:Bursting} (c).
Moreover, starting from the coupling strength $g=30$, where no more collective effects
are present the free spiking period $T_{{\rm free}}$ appears as the minimal $T_{ISI}$ of the distributions 
$P(T_{ISI})$ (see the inset in Fig.~\ref{fig:Bursting} (c)). 

Finally, in the regime of very large $g$, an interesting phenomenon emerges 
shown in Fig.~\ref{fig:Bursting} (d) for $g=1000$: 
the ISI distribution displays a large peak at $T_{{\rm free}}$ 
and and exponential tail, a typical signature of Poissonian firing.
This structure is due to the bursting activity of the neuron (see also Fig.~\ref{fig:fig3} (f)).
Indeed, for this large coupling the firing rate of the pre-synaptic neurons is very low, therefore
the post-synaptic neurons are usually not inhibited and fire with their free spiking period $T_{{\rm free}}$.
However, whenever they receive sporadically inhibitory kicks of large amplitude $g$, the neurons are silent
for the long period necessary to the membrane potential to recover positive values.
Furthermore, the Poissonian nature of the distribution of the kick arrival times, due to
the network sparsness, reflects in the long tail of the $P(T_{ISI})$.
 
\begin{figure}
\centering
\includegraphics[angle=0,width=0.95\linewidth]{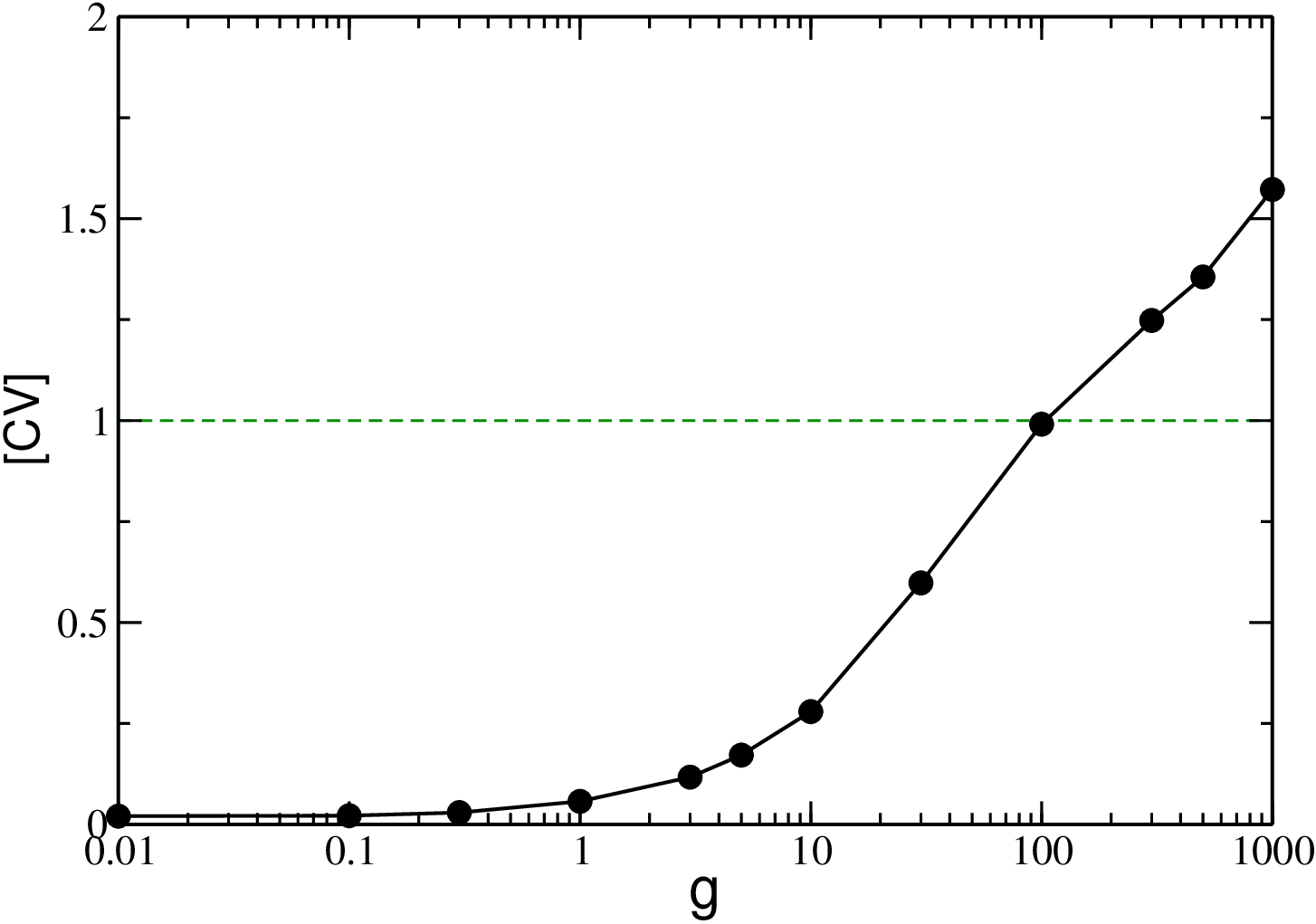}
\caption{\textbf{Microscopic variability of the ISIs in SNs.} 
Ensemble average over all the neurons of the coefficients of variation of the ISIs, $[CV]$,  
versus the coupling strength $g$. The dashed line signals the value corresponding to a 
Poissonian statistics, namely $[CV]=1$. Network size $N=4000$ and $K=240$,
other parameters as in Fig. \ref{fig:fig1}.} 
\label{fig:CV}
\end{figure}

Overall,  upon increasing inhibition, 
on one side we observe that the average frequency of neurons steadily decreases, 
on the other side the neurons tend to fire occasionally
at the fastest possible frequency, namely $1/T_{{\rm free}}$.
This behaviour is joined to a steadily 
increasing variability in the microscopic firing of the neurons,
as clearly shown in Fig.~\ref{fig:CV} where we report 
the ensemble average of the coefficients of variation, $[CV]$.
In particular, we observe upon increasing $g$, a transition from a very regular firing characterized 
by $[CV] \simeq 0$ to a dynamics with $[CV] > 1$, which is a signature of multi-modal ISI distributions.

\subsection{Role of the heterogeneity and of the delay}

\begin{figure}
\includegraphics[angle=0,width=0.95\linewidth]{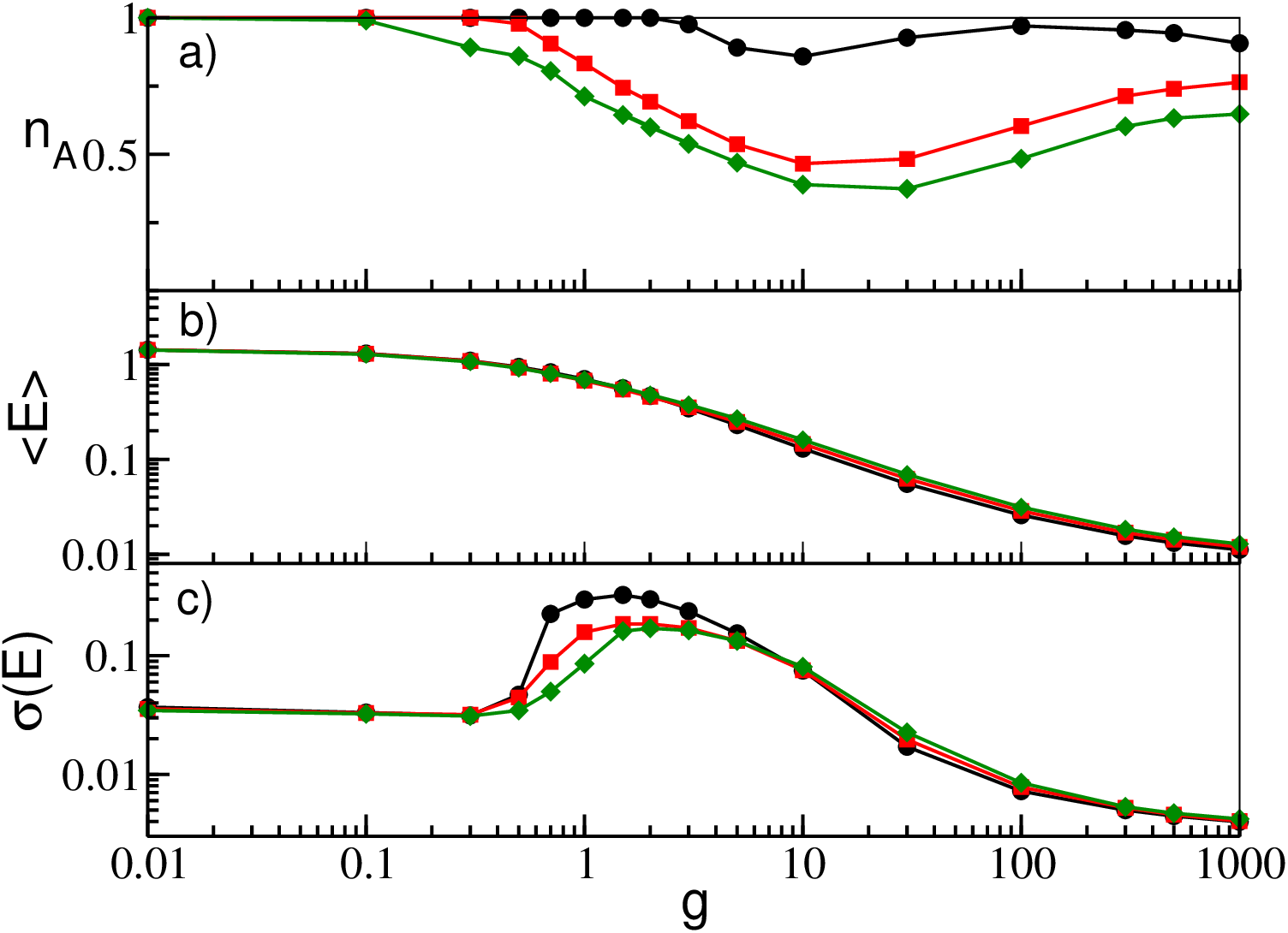}
\caption{\textbf{Relevance of the heterogeneity in SNs.} 
a) Fraction of active neurons $n_A$, b) time average of the mean field $\langle E \rangle$, c) fluctuations 
of the mean field $\sigma(E)$, vs the synaptic inhibition and for several values 
of heterogeneity. Namely, $\Delta a=$0.4 (black circles), $\Delta a=$1.2 (red squares)
$\Delta a=$1.6 (green diamonds). 
In all panels the time delay is set to $t_d = 0.1$, the network size to $N=4000$ and 
the in-degree to $K=240$, other parameters as in Fig. \ref{fig:fig1}.
} 
\label{fig:SparseHeterogeneity}
\end{figure}

\begin{figure}
\includegraphics[angle=0,width=0.95\linewidth]{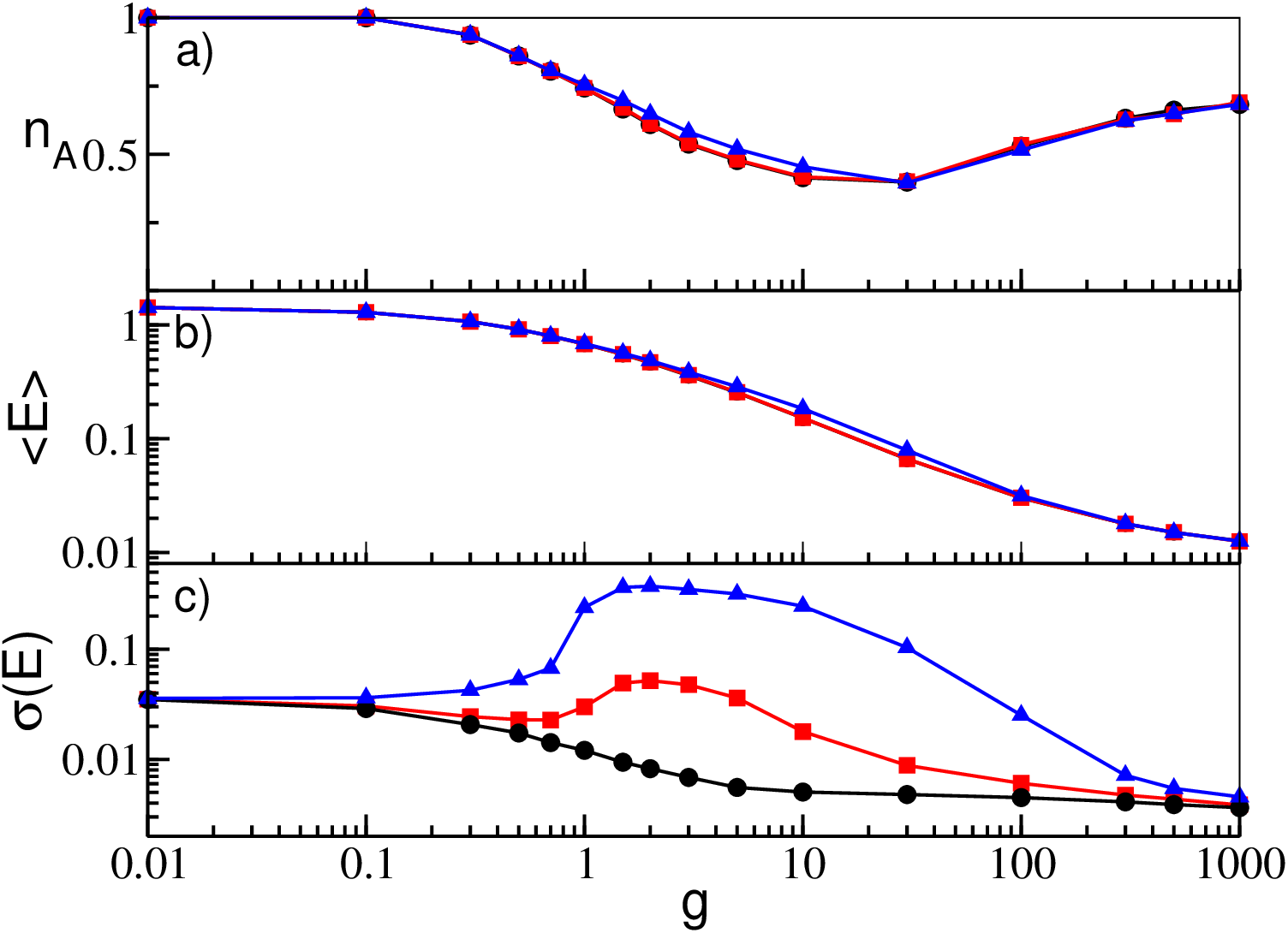}
\caption{\textbf{Relevance of time delay is SNs.} 
a) Fraction of active neurons $n_A$, b) time average of the mean field
$\langle E \rangle$, c) fluctuations 
of the mean field $\sigma(E)$, vs $g$ and different values of 
time delay. Namely, $t_{d}=$0.005 (black circles), $t_{d}=$0.05 (red squares),
and $t_{d}=$0.25 (blue diamonds). 
For this figure a fixed heterogeneity distribution with $\Delta a = 1.6$ is used. 
In all the panels  a fixed value of $N=4000$ and $K = 240$ are employed,
other parameters as in Fig. \ref{fig:fig1}.
} 
\label{fig:SparseDelay}
\end{figure}

Analogously to what done for the GCNs, we  analyzed the influence of different
excitability distributions as well as of the time delay $t_d$ on the dynamics of SNs.
The results of these analyses are reported in Figs. \ref{fig:SparseHeterogeneity}
and \ref{fig:SparseDelay}. 

In order to study the effect of the heterogeneity in the neuronal excitabilities, we choose to 
mantain constant the average $[a]$=2 and to vary $\Delta a$. As discussed in the 
previous sub-sections, heterogeneity is necessary for
the WTA mechanism to kick in. Therefore for small $\Delta a$ the overall
deactivation-reactivation effect is less evident, because the percentage of inactive
neurons is much smaller then at larger $\Delta a$ and the complete reactivation of all neurons
is obtained at relatively smaller $g$ (see Fig. \ref{fig:SparseHeterogeneity} (a)).
Similarly to the GCNs, the average network activity as measured by $\langle E \rangle$ remains unchanged, 
because it is mainly dictated by the average synaptic current (see Fig. \ref{fig:SparseHeterogeneity} (b)).
Finally, also in this case, the value of $\Delta a$ affects 
the onset and the amplitude of the collective motion (see Fig. \ref{fig:SparseHeterogeneity} (c)), 
due to the same mechanism already discussed for GCNs.

Regarding the delay, it is worth to remind that in 
the globally coupled system the effects of the synaptic delays 
were observable for $n_A$ and $\langle E \rangle$ only at large inhibition,
where the WTA mechanism reduces largely the number of active neurons.
This effect is not present in SNs due to the reactivation process 
occurring at sufficiently large $g$ (see Fig. \ref{fig:SparseDelay} (a,b)).
However, similarly to the GCNs, the collective activity can emerge only
for sufficiently long delays, namely $t_d > 0.005$, and
the amplitude of the collective oscillations, measured by $\sigma(E)$ increases
with $t_d$ as shown in Fig. \ref{fig:SparseDelay} (c).

\section{Finite size effects}

In this Section we report a detailed analysis of the effects of the disorder 
on finite size networks. In GCNs the only source of disorder
is associated to the distribution of the excitabilites, while in SNs, 
the disorder is due also to the random distribution of the connections.
In both cases we consider for each system size 10 different network realizations, 
which implies different excitability and connectivity distributions.

\begin{figure}
\begin{center}
\includegraphics[angle=0,width=0.95\linewidth]{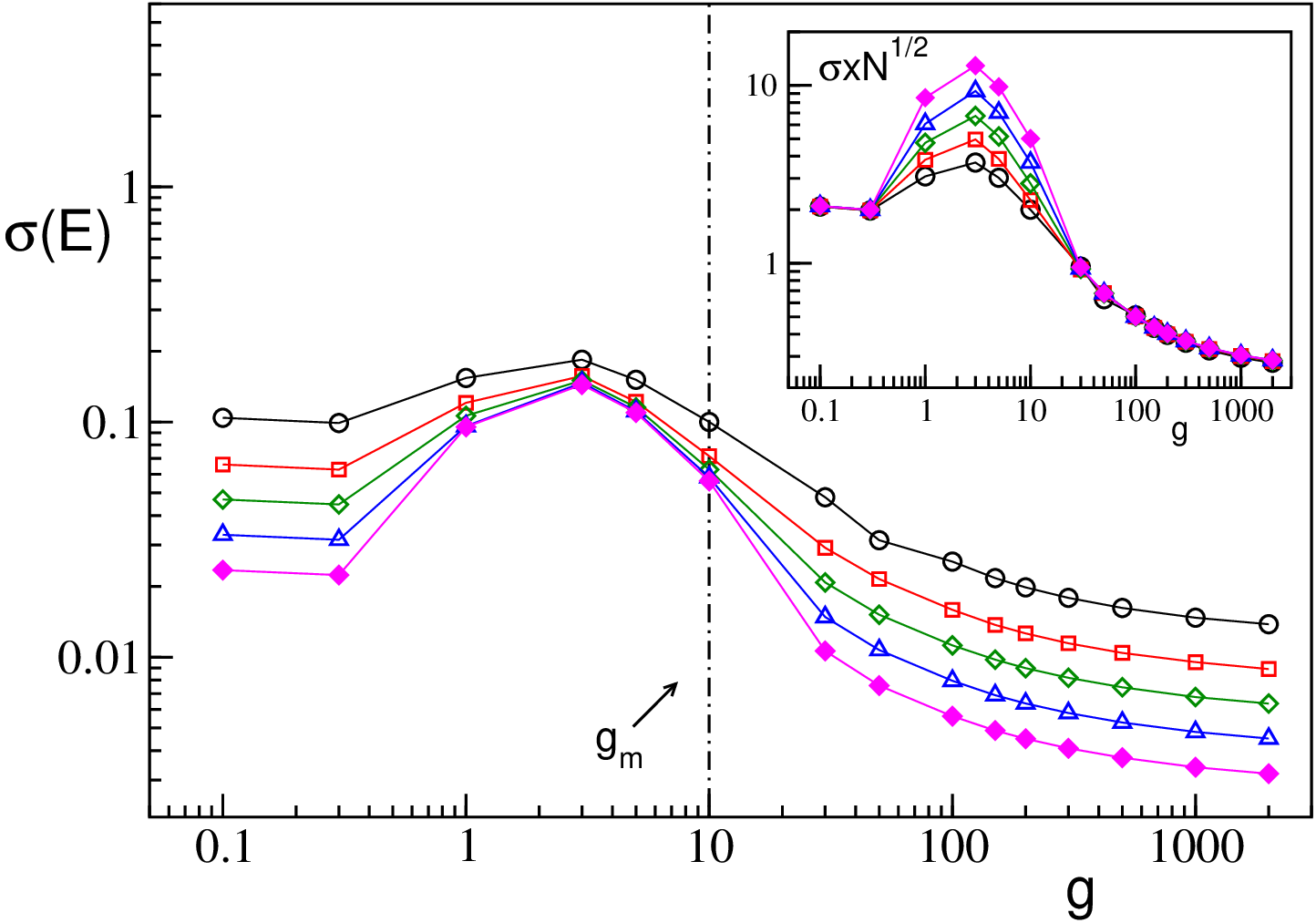}\\
\end{center}
\caption{
\textbf{Finite size scaling of the mean field fluctuations in SNs.} 
Standard deviation of the mean field, $\sigma(E)$, as a function of inhibition for networks with
$K=150$ and increasing size. Namely, $N=400$ (black circles), $N=1000$ (red squares), $N=2000$ (green diamonds), 
$N=4000$ (blue triangles), and $N=8000$ (magenta diamonds). The vertical dot-dashed line marks $g_m$.
In the inset the standard deviation has been rescaled with the system size $N$.  
Each point in the figure and in the inset is the average over 10 realizations of the disorder. 
Other parameters as in Fig. \ref{fig:fig1}.}
\label{fig:FiniteSize}
\end{figure}

Let us first consider the field $E$ and its fluctuations $\sigma(E)$,
similarly to what
reported  for the GCNs (see Fig. \ref{fig:fig1} (b)) 
also for the SNs the average value of $\langle E \rangle$ does not depend on $N$ (data not shown).
Instead, the mean field fluctuations strongly
depend on the size $N$, as shown also for the GCNs in Fig. \ref{fig:fig1} (c).
In particular, for SNs we report $\sigma(E)$ as a function of $g$ 
in Fig. \ref{fig:FiniteSize} for a fixed in-degree $K=150$ 
for system sizes ranging from $N=400$ to $N=8000$.
From the figure (and the inset) it is clear that for $g \le 0.3$ and $g \geq 30$ 
$\sigma(E) \propto N^{-1/2}$ indicating that in the thermodynamic limit the
dynamics is asynchronous for small and large couplings. For intermediate
values of $g$ (namely $ 0.3 < g < 30$) $\sigma(E)$ saturates, for sufficiently
large $N$, to an asymptotic finite value, thus showing clearly the persistence
of the collective behaviour in the thermodynamic limit.
From this analysis 
we can conclude that the collective 
oscillations are present in an interval of $g$ which
remains finite in the thermodynamic limit and 
whose width is determined by the value of $K$
(as shown in Fig.  \ref{fig:fig3} (c)) but not by the size $N$.
Furthermore, for SNs with sufficiently long delay we have two 
phase transitions: from asynchronous to collective behaviour (at small coupling)
and from collective to asynchronous dynamics (at large $g$).

\begin{figure*}[ht]
\begin{center}
\includegraphics*[angle=0,width=8.0cm]{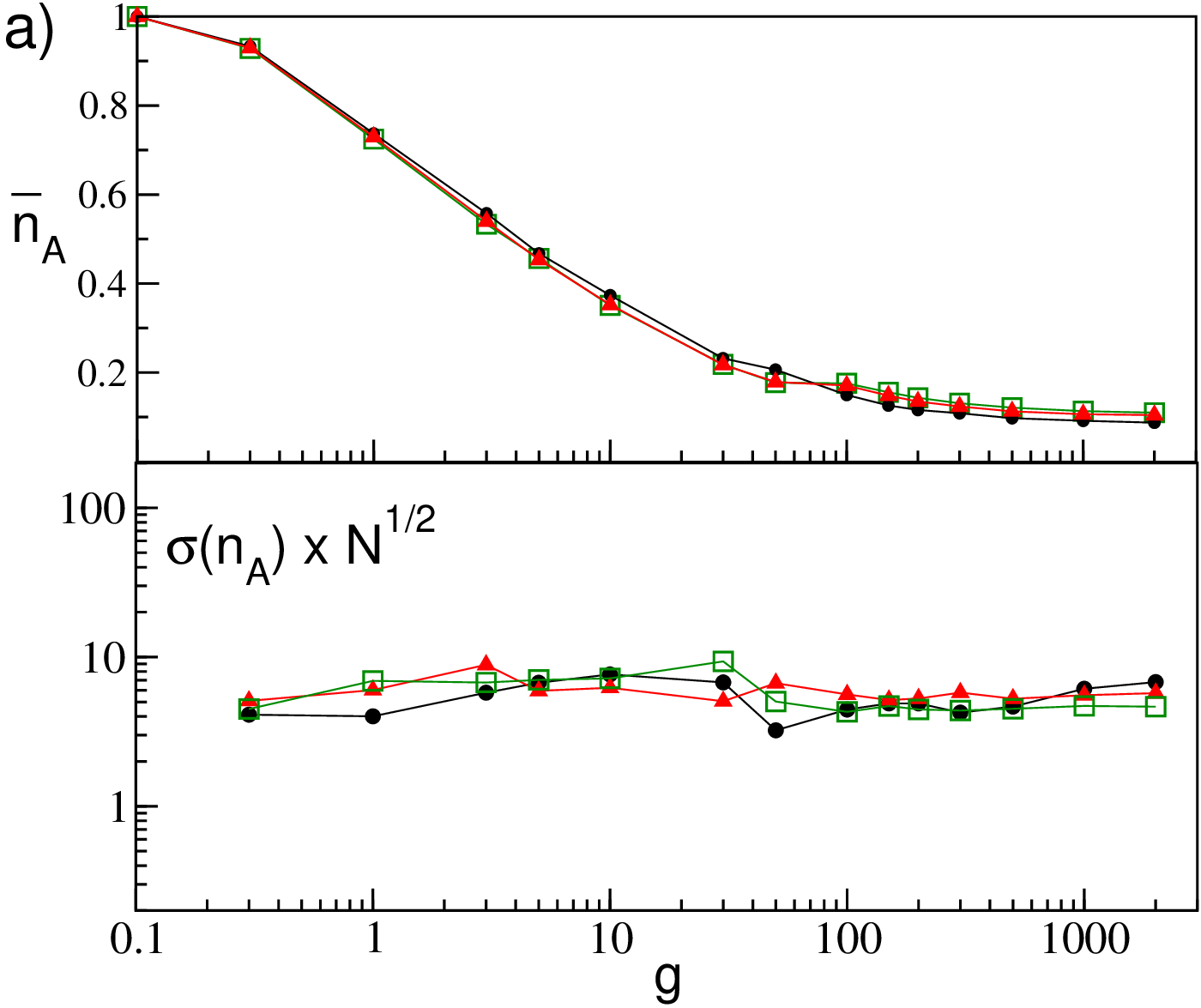}
\includegraphics*[angle=0,width=8.0cm]{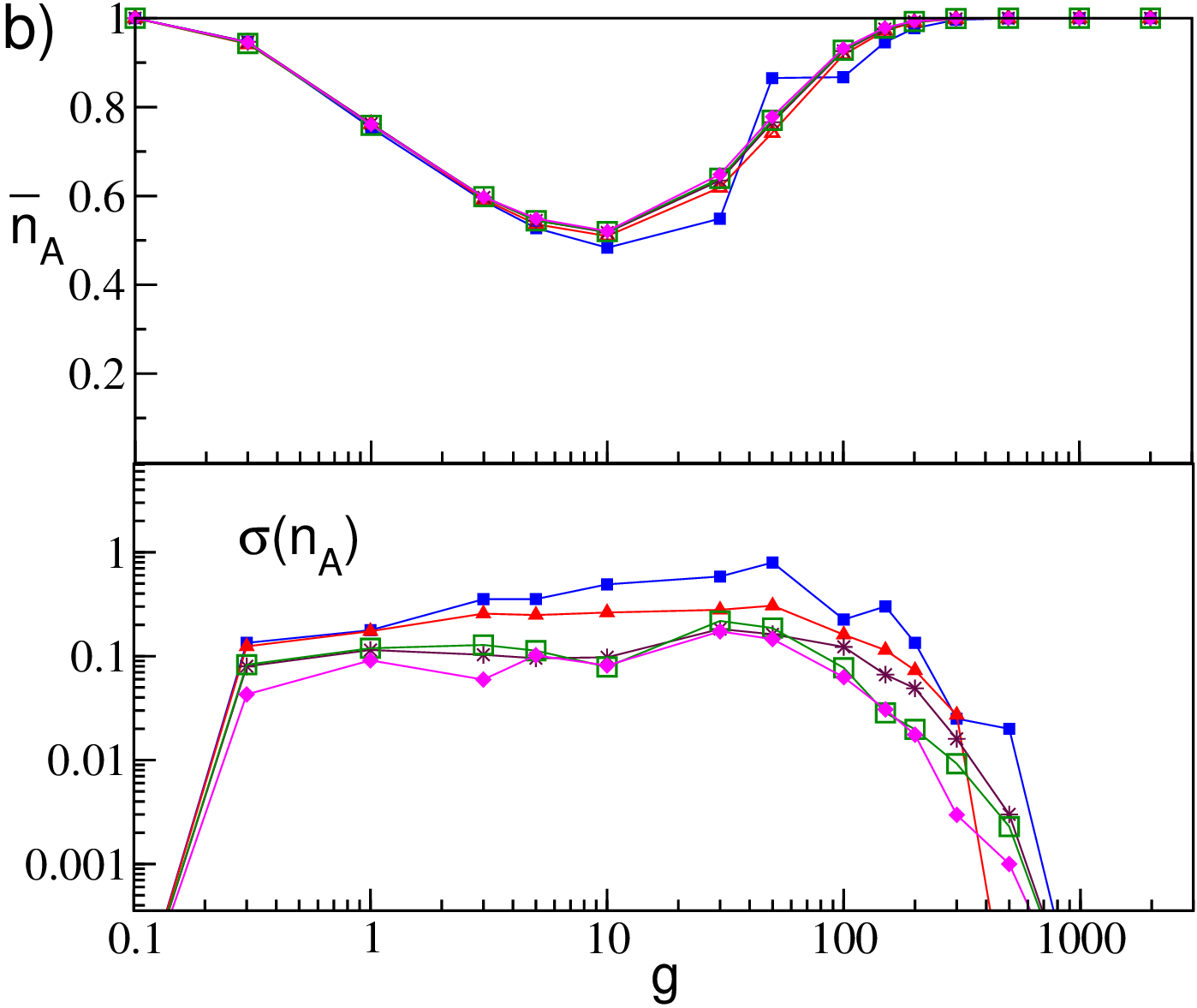}
\end{center}
\caption{{\bf Fraction of active neurons: finite size effects.} 
Panel (a) refers to GCNs, while panel (b) to SNSs with $K=150$, for different network sizes: 
namely, $N=200$ (black circles), $N=400$ (blue squares), $N=1000$ (red triangles), 
$N=2000$ (maroon stars), $N=4000$ (green squares), and $N=8000$ (magenta diamonds). 
The panels display the average $\overline{n_{A}}$ (top) and the standard deviation $\sigma(n_A)$ 
(bottom) of the fraction of active neurons versus the coupling strength $g$. The average and fluctuations
have been measured over 10 different realizations 
of the disorder in the network for each value of $g$ and $N$. In the case of GCNs the standard 
deviation has been multiplied by the square root of the system size. 
Remaining parameters as in Fig.~\ref{fig:fig1}.  
} 
\label{figdisordine:1}
\end{figure*}

Let us now consider the effect of the realizations of 
disorder on the percentage of active neurons. 
In particular, in Fig.~\ref{figdisordine:1} we report the average, $\overline{n}_{A}$, and the standard 
deviation, $\sigma(n_A)$, of the fraction of active neurons obtained for 10 different
network realizations for increasing $N$ both for GCNs an SNs. As a general remark
we observe that $\overline{n}_{A}$ is not  particularly sensible to the system
size, apart for really small sizes ($N < 500$) in the SN case 
(see the upper panels in Fig.~\ref{figdisordine:1} (a) and (b)). The case of small
network sizes for SNs will be discussed in the following of this Section.

For what concerns $\sigma(n_A)$ for the GCNs we observe, as expected,
a decrease as $N^{-1/2}$ with the system size, as clearly evident
from the lower panel in Fig.~\ref{figdisordine:1} (a).
Moreover we observe 
that $\sigma(n_A)$ is essentially constant over the whole range of the coupling strength, 
apart the case of very small coupling strength, $g\le 0.1$ (not shown in the figure), where 
due to the essentially uncoupled dynamics of the neurons $\sigma(n_A)=0$ for every $N$. 
The behavior is quite different for SNs as it is shown in 
Fig.~\ref{figdisordine:1} (b) for networks with $K=150$. 
As a general remark we observe that whenever $\overline{n}_{A}\rightarrow 1$
(i.e. for $g < 0.3 $ and $g > 1000$) the fluctuations vanish and
$\sigma(n_A)$ exhibit finite values in the range of synaptc strenght
where $\overline{n}_{A} <  1$. Furthermore, for increasing $N$
the values of $\sigma(n_A)$ saturate towards an asymptotic profile.
Therefore the fluctuations will persist even
in the thermodynamic limit, in agreement with the results reported in~\cite{Luccioli2012PRL},
and they assume an almost constant value ($\sigma(n_A) \simeq 0.1$) in the range
of existence of collective oscillations (namely, $0.3 < g < 30$).

\begin{figure*}[ht]
\begin{center}
\includegraphics*[angle=0,width=8.0cm]{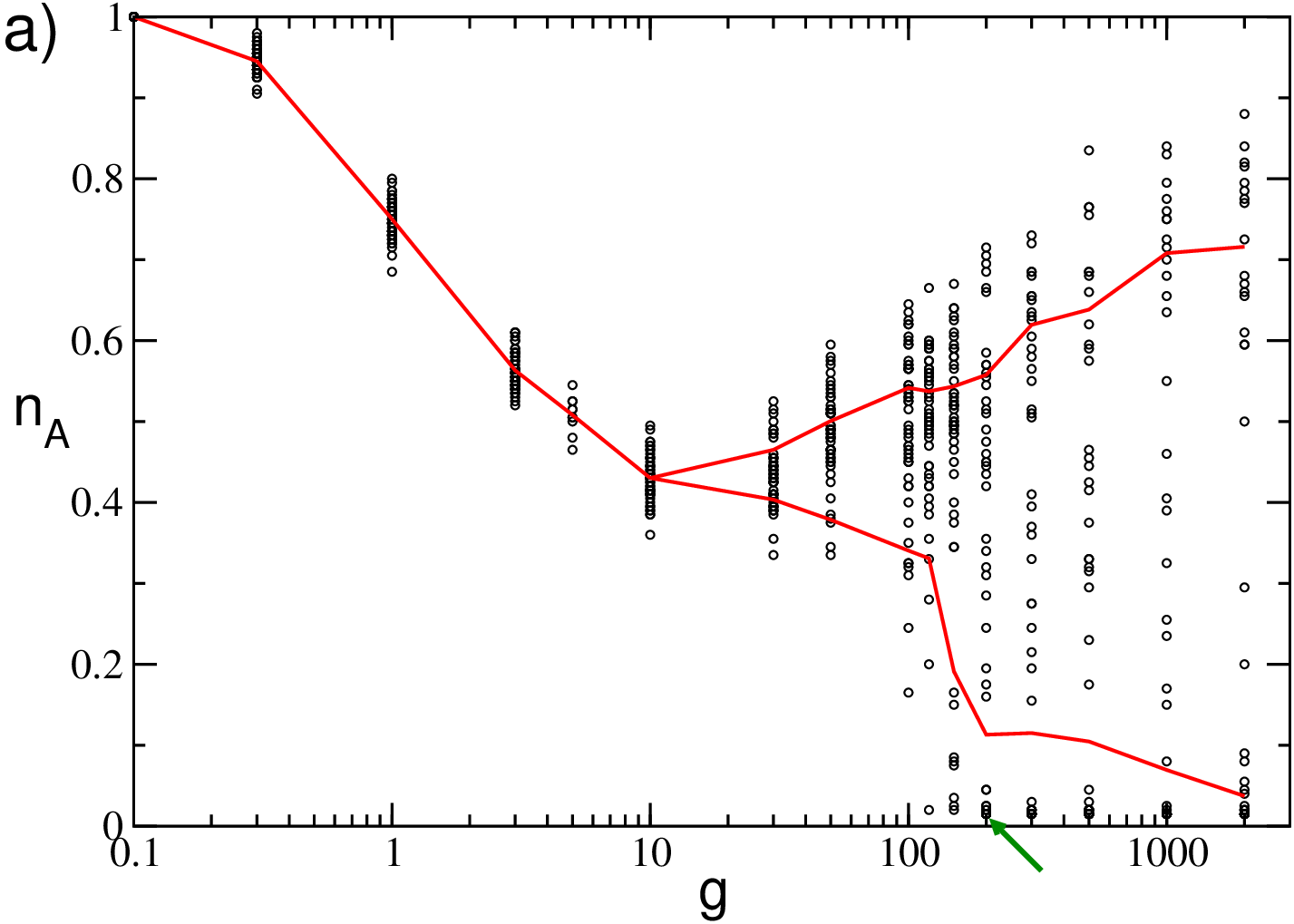}
\includegraphics*[angle=0,width=8.0cm]{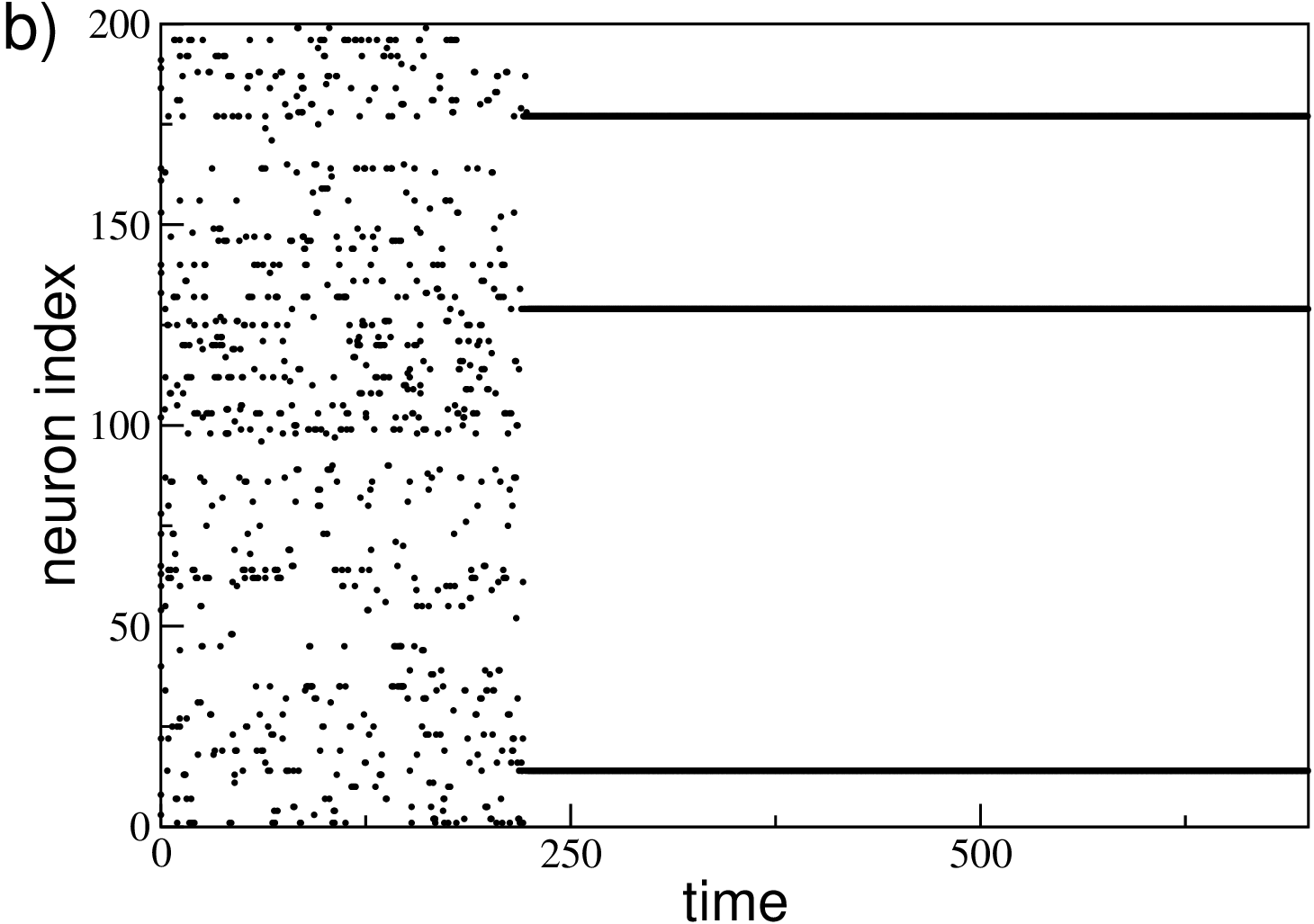}
\end{center}
\caption{{\bf Effects of the realizations of disorder on small SNs.} 
(a) The plot reports the values of $n_A$ obtained for 50 different realizations 
of the network for each value of the coupling strength $g$. 
The red lines are averages ${\bar n}_A$ over the realizations of the disorder: from $0.1 \ge g \le 10$ 
the average is computed over the the total number of realizations, while for coupling strength 
larger than $g=10$ the averages are performed by considering two groups of networks as explained in the text. 
(b) The raster plot displays an example of a the dynamical evolution of a peculiar state obtained for large coupling (here $g=200$), which is indicated in panel (a) by a green arrow. 
Data refer to $N=200$ and $K=150$, other parameters as in Fig.~\ref{fig:fig1}.} 
\label{figdisordine:2}
\end{figure*}

The sparsness of the network can give rise to striking effects for small 
system size, as it is 
shown in Fig.~\ref{figdisordine:2}. 
In particular, in Fig.~\ref{figdisordine:2} (a) we report the values 
of $n_A$ for 50 different realizations 
of the network for $N=200$ and $K=150$. 
For small coupling strength, namely $g < g_m =10$, we observe
that the distribution of $n_A$ values has a single peak centered
around the average ${\bar n}_A$. While, for larger coupling strength the distribution
reveals two distinct peaks: one associated to the typical dynamics 
of a sparse network at large $g$ (i.e. neural reactivation) 
and one to the typical dynamics of a GCN (i.e. the WTA mechanism). 
Thus rendering the definition of ${\bar n}_A$ quite
questionable. As a matter of fact, for $g > g_m =10$ we estimated two 
distinct averages for each $g$, one based on the $n_A$ values larger
than $n_m  \equiv {\bar n}_A(g_m)=0.43$ and one on the smaller
values, these are reported as red lines in Fig.~\ref{figdisordine:2} (a).
We observe this coexistence of two different type of dynamics also 
by considering  different initial conditions for a fixed disorder realization (data not shown).

A peculiar dynamical state can be observed for sufficiently large $g > 100$,
denoted by a green arrow in Fig.~\ref{figdisordine:2} (a).
In this case the corresponding raster plot (shown in Fig.~\ref{figdisordine:2} (b)) 
reveals, after a short transient, the convergence towards a dynamical state where
only few neurons survive (namely, three in this case), while the rest of the network
becomes silent. The interesting aspect is that these three neurons are completely uncoupled
among them and their activity is sufficient to silence all the rest of the neurons.
The microscopic analysis reveals that the three neurons have high intrinsic 
excitability (but not the highest) and that the ensemble of their
post-synaptic neurons correspond to the whole network, apart themselves.

The reported effects, i.e. the coexistence of different dynamics as well as  
the existence of states made of totally uncoupled neurons, disappear increasing 
the system size. As a matter of fact these effects are already no more observable for $N=400$.

\section{Concluding remarks}

In this paper we have clarified how in an inhibitory spiking network
the introduction of various ingredients, characteristic of real brain circuits, like
delay in the electric signal transmission, heterogeneity of the
neurons and random sparseness in the synaptic connections,
can influence the neural dynamics.

In particular, we have studied at a macroscopic and microscopic level the
dynamics of heterogeneous inhibitory spiking networks with delay
for increasing synaptic coupling.
In GCNs the heterogeneity is responsible for neuron's death 
via the WTA mechanism, while the delay allows for the emergence
of COs beyond a critical coupling strength. Furthermore, we
have shown that the increase in the delay favours the overall
collective dynamics (synchronization) in the system, thus reducing the effective variability in the neuron dynamics. Therefore, longer delays counteract the
effect of heterogeneity in the system, which promotes neural
deactivation and asynchronous dynamics.

In SNs by increasing the coupling we observe a passage
from a mean driven to a fluctuation driven dynamics
induced by the sparsness in the connections.
We have a transition from a regime where the neurons are on
average supra-threshold to a phase where they are on average
below-threshold and their firing is induced by large fluctuations in
the currents. This transition is signaled by the occurrence of
a minimum in the value of the fraction of active neurons as a 
function of the inhibitory coupling. Therefore we can affirm that
we pass from a regime dominated by the WTA mechanism, to an activation 
regime controlled by fluctuations, where all neurons are finally firing but
with firing rates definitely lower then those dictated by their
excitabilities \cite{angulo2017death_NJP}.
However, the fluctuations desynchronize the neural dynamics:
the COs emerging at small coupling, due to the time
delay, disappear at large coupling when current fluctuations 
become predominant in the neural dynamics. 

Finite size analysis confirm that in SNs we have two 
phase transitions that delimit the finite range of
couplings where COs are observable.
Outside this range the dynamics is asynchronous,
however we have two different kinds of asynchronous dynamics at low and
high coupling.
At low coupling, we observe a situation where the firing variability of
each neuron is quite low and essentially the active neurons behave
almost independently. At large coupling, the variability of the firing
activity is extremely large, characterized by a bursting behaviour at the
level of single neurons. Due to the sparsness and the low activity 
of the fluctuations activated pre-synaptic neurons, each
neuron is subject to low rate Poissonian spike trains of PSPs of large amplitude.
Therefore the neurons are for long time active and unaffected by 
the other neurons, however when they receive large inhibitory PSPs 
they remain silent for long periods.

It has been shown that heterogeneity and noise can increase the information
encoded by a population counteracting the correlation present in neuronal activity
\cite{shamir2006,padmanabhan2010,ecker2011,mejias2014,mejias2012Optimal}. 
However, it remains to clarify how disorder (neural heterogeneity 
and randomness in the connections) and delay should combine 
to enhance information encoding. The results reported in this paper 
can help in understanding the influence of delay and disorder on the dynamics of
neural circuits and therefore on their ability to store and recover information.

\acknowledgements
The authors acknowledge N. Brunel, V. Hakim, S. Olmi, for enlightening discussions. 
AT has received partial support by the Excellence Initiative I-Site Paris Seine (No ANR-16-IDEX-008), 
by the Labex MME-DII (No ANR-11-LBX-0023-01) and by the ANR Project ERMUNDY (No ANR-18-CE37-0014) 
all funded by the French programme ``Investissements d'Avenir''. DA-G has received financial support from
Vicerrectoria de Investigaciones - Universidad de Cartagena (Project No. 085-2018) and from
CNRS for a research period at LPTM,  UMR 8089, Universit\'e de Cergy-Pontoise, France.
The work has been partly realized at the Max Planck Institute for the Physics of Complex Systems (Dresden, Germany) as part of the activity 
of the Advanced Study Group 2016/17 ``From Microscopic to Collective Dynamics in Neural Circuits''.

%%\bibliographystyle{apsrev-nourl}
%%\bibliography{Bibliography_Rebirth}
%%\bibliography{Bibliography_report}
%%\bibliography{//home/dangulog/Dropbox/ANGULO/BibliographyComplete/Bibliography_report}

\begin{thebibliography}{39}
\expandafter\ifx\csname natexlab\endcsname\relax\def\natexlab#1{#1}\fi
\expandafter\ifx\csname bibnamefont\endcsname\relax
  \def\bibnamefont#1{#1}\fi
\expandafter\ifx\csname bibfnamefont\endcsname\relax
  \def\bibfnamefont#1{#1}\fi
\expandafter\ifx\csname citenamefont\endcsname\relax
  \def\citenamefont#1{#1}\fi
\expandafter\ifx\csname url\endcsname\relax
  \def\url#1{\texttt{#1}}\fi
\expandafter\ifx\csname urlprefix\endcsname\relax\def\urlprefix{URL }\fi
\providecommand{\bibinfo}[2]{#2}
\providecommand{\eprint}[2][]{\url{#2}}

\bibitem[{\citenamefont{Ben-Ari}(2001)}]{ben2001developing}
\bibinfo{author}{\bibfnamefont{Y.}~\bibnamefont{Ben-Ari}},
  \bibinfo{journal}{Trends in neurosciences} \textbf{\bibinfo{volume}{24}},
  \bibinfo{pages}{353} (\bibinfo{year}{2001}).

\bibitem[{\citenamefont{Jonas and Buzsaki}(2007)}]{jonas2007}
\bibinfo{author}{\bibfnamefont{P.}~\bibnamefont{Jonas}} \bibnamefont{and}
  \bibinfo{author}{\bibfnamefont{G.}~\bibnamefont{Buzsaki}},
  \bibinfo{journal}{Scholarpedia} \textbf{\bibinfo{volume}{2}},
  \bibinfo{pages}{3286} (\bibinfo{year}{2007}).

\bibitem[{\citenamefont{Whittington et~al.}(2000)\citenamefont{Whittington,
  Traub, Kopell, Ermentrout, and Buhl}}]{whittington2000}
\bibinfo{author}{\bibfnamefont{M.~A.} \bibnamefont{Whittington}},
  \bibinfo{author}{\bibfnamefont{R.}~\bibnamefont{Traub}},
  \bibinfo{author}{\bibfnamefont{N.}~\bibnamefont{Kopell}},
  \bibinfo{author}{\bibfnamefont{B.}~\bibnamefont{Ermentrout}},
  \bibnamefont{and} \bibinfo{author}{\bibfnamefont{E.}~\bibnamefont{Buhl}},
  \bibinfo{journal}{International journal of psychophysiology}
  \textbf{\bibinfo{volume}{38}}, \bibinfo{pages}{315} (\bibinfo{year}{2000}).

\bibitem[{\citenamefont{Buzsaki}(2006)}]{Buzsaki2006Rhythms}
\bibinfo{author}{\bibfnamefont{G.}~\bibnamefont{Buzsaki}},
  \emph{\bibinfo{title}{{Rhythms of the Brain}}} (\bibinfo{publisher}{Oxford
  University Press, USA}, \bibinfo{year}{2006}), \bibinfo{edition}{1st} ed.,
  ISBN \bibinfo{isbn}{0195301064}.

\bibitem[{\citenamefont{Buzs{\'a}ki and Chrobak}(1995)}]{buzsaki1995}
\bibinfo{author}{\bibfnamefont{G.}~\bibnamefont{Buzs{\'a}ki}} \bibnamefont{and}
  \bibinfo{author}{\bibfnamefont{J.~J.} \bibnamefont{Chrobak}},
  \bibinfo{journal}{Current opinion in neurobiology}
  \textbf{\bibinfo{volume}{5}}, \bibinfo{pages}{504} (\bibinfo{year}{1995}).

\bibitem[{\citenamefont{Salinas and Sejnowski}(2001)}]{salinas2001}
\bibinfo{author}{\bibfnamefont{E.}~\bibnamefont{Salinas}} \bibnamefont{and}
  \bibinfo{author}{\bibfnamefont{T.~J.} \bibnamefont{Sejnowski}},
  \bibinfo{journal}{Nature reviews neuroscience} \textbf{\bibinfo{volume}{2}},
  \bibinfo{pages}{539} (\bibinfo{year}{2001}).

\bibitem[{\citenamefont{van Vreeswijk}(1996)}]{vanVreeswijk1996Partial}
\bibinfo{author}{\bibfnamefont{C.}~\bibnamefont{van Vreeswijk}},
  \bibinfo{journal}{Phys. Rev. E} \textbf{\bibinfo{volume}{54}},
  \bibinfo{pages}{5522} (\bibinfo{year}{1996}).

\bibitem[{\citenamefont{Ernst et~al.}(1995)\citenamefont{Ernst, Pawelzik, and
  Geisel}}]{ernst1995}
\bibinfo{author}{\bibfnamefont{U.}~\bibnamefont{Ernst}},
  \bibinfo{author}{\bibfnamefont{K.}~\bibnamefont{Pawelzik}}, \bibnamefont{and}
  \bibinfo{author}{\bibfnamefont{T.}~\bibnamefont{Geisel}},
  \bibinfo{journal}{Physical review letters} \textbf{\bibinfo{volume}{74}},
  \bibinfo{pages}{1570} (\bibinfo{year}{1995}).

\bibitem[{\citenamefont{Friedrich and Kinzel}(2009)}]{kinzel2009}
\bibinfo{author}{\bibfnamefont{J.}~\bibnamefont{Friedrich}} \bibnamefont{and}
  \bibinfo{author}{\bibfnamefont{W.}~\bibnamefont{Kinzel}},
  \bibinfo{journal}{Journal of computational neuroscience}
  \textbf{\bibinfo{volume}{27}}, \bibinfo{pages}{65} (\bibinfo{year}{2009}).

\bibitem[{\citenamefont{Zillmer et~al.}(2006)\citenamefont{Zillmer, Livi,
  Politi, and Torcini}}]{Zillmer2006}
\bibinfo{author}{\bibfnamefont{R.}~\bibnamefont{Zillmer}},
  \bibinfo{author}{\bibfnamefont{R.}~\bibnamefont{Livi}},
  \bibinfo{author}{\bibfnamefont{A.}~\bibnamefont{Politi}}, \bibnamefont{and}
  \bibinfo{author}{\bibfnamefont{A.}~\bibnamefont{Torcini}},
  \bibinfo{journal}{Phys. Rev. E} \textbf{\bibinfo{volume}{74}},
  \bibinfo{pages}{036203} (\bibinfo{year}{2006}).

\bibitem[{\citenamefont{Brunel and Hakim}(1999)}]{BrunelHakim1999}
\bibinfo{author}{\bibfnamefont{N.}~\bibnamefont{Brunel}} \bibnamefont{and}
  \bibinfo{author}{\bibfnamefont{V.}~\bibnamefont{Hakim}},
  \bibinfo{journal}{Neural. Comput.} \textbf{\bibinfo{volume}{11}},
  \bibinfo{pages}{1621} (\bibinfo{year}{1999}).

\bibitem[{\citenamefont{Brunel}(2000)}]{Brunel2000Sparse}
\bibinfo{author}{\bibfnamefont{N.}~\bibnamefont{Brunel}}, \bibinfo{journal}{J.
  Comput. Neurosci.} \textbf{\bibinfo{volume}{8}}, \bibinfo{pages}{183}
  (\bibinfo{year}{2000}), ISSN \bibinfo{issn}{0929-5313}.

\bibitem[{\citenamefont{Politi and Luccioli}(2010)}]{politi2010lif}
\bibinfo{author}{\bibfnamefont{A.}~\bibnamefont{Politi}} \bibnamefont{and}
  \bibinfo{author}{\bibfnamefont{S.}~\bibnamefont{Luccioli}}, in
  \emph{\bibinfo{booktitle}{Network Science: Complexity in Nature and
  Technology}} (\bibinfo{publisher}{Springer, London}, \bibinfo{year}{2010}),
  p. \bibinfo{pages}{217}.

\bibitem[{\citenamefont{Golomb and Hansel}(2000)}]{golomb2000number}
\bibinfo{author}{\bibfnamefont{D.}~\bibnamefont{Golomb}} \bibnamefont{and}
  \bibinfo{author}{\bibfnamefont{D.}~\bibnamefont{Hansel}},
  \bibinfo{journal}{Neural computation} \textbf{\bibinfo{volume}{12}},
  \bibinfo{pages}{1095} (\bibinfo{year}{2000}).

\bibitem[{\citenamefont{di~Volo and Torcini}(2018)}]{di2018}
\bibinfo{author}{\bibfnamefont{M.}~\bibnamefont{di~Volo}} \bibnamefont{and}
  \bibinfo{author}{\bibfnamefont{A.}~\bibnamefont{Torcini}},
  \bibinfo{journal}{Physical review letters} \textbf{\bibinfo{volume}{121}},
  \bibinfo{pages}{128301} (\bibinfo{year}{2018}).

\bibitem[{\citenamefont{Coultrip et~al.}(1992)\citenamefont{Coultrip, Granger,
  and Lynch}}]{coultrip1992cortical}
\bibinfo{author}{\bibfnamefont{R.}~\bibnamefont{Coultrip}},
  \bibinfo{author}{\bibfnamefont{R.}~\bibnamefont{Granger}}, \bibnamefont{and}
  \bibinfo{author}{\bibfnamefont{G.}~\bibnamefont{Lynch}},
  \bibinfo{journal}{Neural networks} \textbf{\bibinfo{volume}{5}},
  \bibinfo{pages}{47} (\bibinfo{year}{1992}).

\bibitem[{\citenamefont{Fukai and Tanaka}(1997)}]{fukai1997simple}
\bibinfo{author}{\bibfnamefont{T.}~\bibnamefont{Fukai}} \bibnamefont{and}
  \bibinfo{author}{\bibfnamefont{S.}~\bibnamefont{Tanaka}},
  \bibinfo{journal}{Neural computation} \textbf{\bibinfo{volume}{9}},
  \bibinfo{pages}{77} (\bibinfo{year}{1997}).

\bibitem[{\citenamefont{Itti and Koch}(2001)}]{itti2001}
\bibinfo{author}{\bibfnamefont{L.}~\bibnamefont{Itti}} \bibnamefont{and}
  \bibinfo{author}{\bibfnamefont{C.}~\bibnamefont{Koch}},
  \bibinfo{journal}{Nature reviews neuroscience} \textbf{\bibinfo{volume}{2}},
  \bibinfo{pages}{194} (\bibinfo{year}{2001}).

\bibitem[{\citenamefont{Lee et~al.}(1999)\citenamefont{Lee, Itti, Koch, and
  Braun}}]{lee1999}
\bibinfo{author}{\bibfnamefont{D.~K.} \bibnamefont{Lee}},
  \bibinfo{author}{\bibfnamefont{L.}~\bibnamefont{Itti}},
  \bibinfo{author}{\bibfnamefont{C.}~\bibnamefont{Koch}}, \bibnamefont{and}
  \bibinfo{author}{\bibfnamefont{J.}~\bibnamefont{Braun}},
  \bibinfo{journal}{Nature neuroscience} \textbf{\bibinfo{volume}{2}},
  \bibinfo{pages}{375} (\bibinfo{year}{1999}).

\bibitem[{\citenamefont{Wang}(2002)}]{wang2002probabilistic}
\bibinfo{author}{\bibfnamefont{X.-J.} \bibnamefont{Wang}},
  \bibinfo{journal}{Neuron} \textbf{\bibinfo{volume}{36}}, \bibinfo{pages}{955}
  (\bibinfo{year}{2002}).

\bibitem[{\citenamefont{Wong and Wang}(2006)}]{wong2006recurrent}
\bibinfo{author}{\bibfnamefont{K.-F.} \bibnamefont{Wong}} \bibnamefont{and}
  \bibinfo{author}{\bibfnamefont{X.-J.} \bibnamefont{Wang}},
  \bibinfo{journal}{Journal of Neuroscience} \textbf{\bibinfo{volume}{26}},
  \bibinfo{pages}{1314} (\bibinfo{year}{2006}).

\bibitem[{\citenamefont{Fries et~al.}(2007)\citenamefont{Fries, Nikoli{\'c},
  and Singer}}]{fries2007}
\bibinfo{author}{\bibfnamefont{P.}~\bibnamefont{Fries}},
  \bibinfo{author}{\bibfnamefont{D.}~\bibnamefont{Nikoli{\'c}}},
  \bibnamefont{and} \bibinfo{author}{\bibfnamefont{W.}~\bibnamefont{Singer}},
  \bibinfo{journal}{Trends in neurosciences} \textbf{\bibinfo{volume}{30}},
  \bibinfo{pages}{309} (\bibinfo{year}{2007}).

\bibitem[{\citenamefont{Angulo-Garcia et~al.}(2017)\citenamefont{Angulo-Garcia,
  Luccioli, Olmi, and Torcini}}]{angulo2017death_NJP}
\bibinfo{author}{\bibfnamefont{D.}~\bibnamefont{Angulo-Garcia}},
  \bibinfo{author}{\bibfnamefont{S.}~\bibnamefont{Luccioli}},
  \bibinfo{author}{\bibfnamefont{S.}~\bibnamefont{Olmi}}, \bibnamefont{and}
  \bibinfo{author}{\bibfnamefont{A.}~\bibnamefont{Torcini}},
  \bibinfo{journal}{New Journal of Physics} \textbf{\bibinfo{volume}{19}},
  \bibinfo{pages}{053011} (\bibinfo{year}{2017}).

\bibitem[{\citenamefont{Ponzi and Wickens}(2013)}]{ponzi2013optimal}
\bibinfo{author}{\bibfnamefont{A.}~\bibnamefont{Ponzi}} \bibnamefont{and}
  \bibinfo{author}{\bibfnamefont{J.~R.} \bibnamefont{Wickens}},
  \bibinfo{journal}{PLoS computational biology} \textbf{\bibinfo{volume}{9}},
  \bibinfo{pages}{e1002954} (\bibinfo{year}{2013}).

\bibitem[{\citenamefont{Angulo-Garcia et~al.}(2015)\citenamefont{Angulo-Garcia,
  Berke, and Torcini}}]{angulo2015Striatum}
\bibinfo{author}{\bibfnamefont{D.}~\bibnamefont{Angulo-Garcia}},
  \bibinfo{author}{\bibfnamefont{J.~D.} \bibnamefont{Berke}}, \bibnamefont{and}
  \bibinfo{author}{\bibfnamefont{A.}~\bibnamefont{Torcini}},
  \bibinfo{journal}{PLoS Comput Biol} \textbf{\bibinfo{volume}{12}},
  \bibinfo{pages}{e1004778} (\bibinfo{year}{2015}).

\bibitem[{\citenamefont{Olmi et~al.}(2012)\citenamefont{Olmi, Politi, and
  Torcini}}]{olmi2012}
\bibinfo{author}{\bibfnamefont{S.}~\bibnamefont{Olmi}},
  \bibinfo{author}{\bibfnamefont{A.}~\bibnamefont{Politi}}, \bibnamefont{and}
  \bibinfo{author}{\bibfnamefont{A.}~\bibnamefont{Torcini}},
  \bibinfo{journal}{The Journal of Mathematical Neuroscience}
  \textbf{\bibinfo{volume}{2}}, \bibinfo{pages}{12} (\bibinfo{year}{2012}).

\bibitem[{\citenamefont{Luccioli and Politi}(2010)}]{Luccioli2010Irregular}
\bibinfo{author}{\bibfnamefont{S.}~\bibnamefont{Luccioli}} \bibnamefont{and}
  \bibinfo{author}{\bibfnamefont{A.}~\bibnamefont{Politi}},
  \bibinfo{journal}{Phys. Rev. Lett.} \textbf{\bibinfo{volume}{105}},
  \bibinfo{pages}{158104+} (\bibinfo{year}{2010}).

\bibitem[{\citenamefont{Zillmer et~al.}(2009)\citenamefont{Zillmer, Brunel, and
  Hansel}}]{Zillmer2009LongTrans}
\bibinfo{author}{\bibfnamefont{R.}~\bibnamefont{Zillmer}},
  \bibinfo{author}{\bibfnamefont{N.}~\bibnamefont{Brunel}}, \bibnamefont{and}
  \bibinfo{author}{\bibfnamefont{D.}~\bibnamefont{Hansel}},
  \bibinfo{journal}{Phys. Rev. E} \textbf{\bibinfo{volume}{79}},
  \bibinfo{pages}{031909} (\bibinfo{year}{2009}).

\bibitem[{\citenamefont{Politi and Torcini}(2010)}]{politi2010stable}
\bibinfo{author}{\bibfnamefont{A.}~\bibnamefont{Politi}} \bibnamefont{and}
  \bibinfo{author}{\bibfnamefont{A.}~\bibnamefont{Torcini}}, in
  \emph{\bibinfo{booktitle}{Nonlinear Dynamics and Chaos: Advances and
  Perspectives}} (\bibinfo{publisher}{Springer}, \bibinfo{year}{2010}), pp.
  \bibinfo{pages}{103--129}.

\bibitem[{not()}]{note1}
\bibinfo{note}{The {\it winners} are the neurons with higher values of the
  excitability. In the case of a uniform distribution defined in the interval
  $[a_1 : a_2]$ the smallest excitability of an active neuron is $a_S =a_2 -
  n_A \Delta a$. Therefore the spread of the excitabilities of the ative neuron
  is given by $a_2 - a_S = n_A \Delta a$.}

\bibitem[{\citenamefont{Ernst et~al.}(1998)\citenamefont{Ernst, Pawelzik, and
  Geisel}}]{ernst1998}
\bibinfo{author}{\bibfnamefont{U.}~\bibnamefont{Ernst}},
  \bibinfo{author}{\bibfnamefont{K.}~\bibnamefont{Pawelzik}}, \bibnamefont{and}
  \bibinfo{author}{\bibfnamefont{T.}~\bibnamefont{Geisel}},
  \bibinfo{journal}{Phys. Rev. E} \textbf{\bibinfo{volume}{57}},
  \bibinfo{pages}{2150} (\bibinfo{year}{1998}).

\bibitem[{\citenamefont{Luccioli et~al.}(2012)\citenamefont{Luccioli, Olmi,
  Politi, and Torcini}}]{Luccioli2012PRL}
\bibinfo{author}{\bibfnamefont{S.}~\bibnamefont{Luccioli}},
  \bibinfo{author}{\bibfnamefont{S.}~\bibnamefont{Olmi}},
  \bibinfo{author}{\bibfnamefont{A.}~\bibnamefont{Politi}}, \bibnamefont{and}
  \bibinfo{author}{\bibfnamefont{A.}~\bibnamefont{Torcini}},
  \bibinfo{journal}{Phys. Rev. Lett.} \textbf{\bibinfo{volume}{109}},
  \bibinfo{pages}{138103} (\bibinfo{year}{2012}).

\bibitem[{\citenamefont{Renart et~al.}(2007)\citenamefont{Renart, Moreno-Bote,
  Wang, and Parga}}]{renart2007}
\bibinfo{author}{\bibfnamefont{A.}~\bibnamefont{Renart}},
  \bibinfo{author}{\bibfnamefont{R.}~\bibnamefont{Moreno-Bote}},
  \bibinfo{author}{\bibfnamefont{X.-J.} \bibnamefont{Wang}}, \bibnamefont{and}
  \bibinfo{author}{\bibfnamefont{N.}~\bibnamefont{Parga}},
  \bibinfo{journal}{Neural. Comput.} \textbf{\bibinfo{volume}{19}},
  \bibinfo{pages}{1} (\bibinfo{year}{2007}).

\bibitem[{\citenamefont{Tuckwell}(2005)}]{tuckwell2005}
\bibinfo{author}{\bibfnamefont{H.~C.} \bibnamefont{Tuckwell}},
  \emph{\bibinfo{title}{Introduction to theoretical neurobiology: Volume 2,
  nonlinear and stochastic theories}}, vol.~\bibinfo{volume}{8}
  (\bibinfo{publisher}{Cambridge University Press}, \bibinfo{year}{2005}).

\bibitem[{\citenamefont{Shamir and Sompolinsky}(2006)}]{shamir2006}
\bibinfo{author}{\bibfnamefont{M.}~\bibnamefont{Shamir}} \bibnamefont{and}
  \bibinfo{author}{\bibfnamefont{H.}~\bibnamefont{Sompolinsky}},
  \bibinfo{journal}{Neural computation} \textbf{\bibinfo{volume}{18}},
  \bibinfo{pages}{1951} (\bibinfo{year}{2006}).

\bibitem[{\citenamefont{Padmanabhan and Urban}(2010)}]{padmanabhan2010}
\bibinfo{author}{\bibfnamefont{K.}~\bibnamefont{Padmanabhan}} \bibnamefont{and}
  \bibinfo{author}{\bibfnamefont{N.~N.} \bibnamefont{Urban}},
  \bibinfo{journal}{Nature neuroscience} \textbf{\bibinfo{volume}{13}},
  \bibinfo{pages}{1276} (\bibinfo{year}{2010}).

\bibitem[{\citenamefont{Ecker et~al.}(2011)\citenamefont{Ecker, Berens, Tolias,
  and Bethge}}]{ecker2011}
\bibinfo{author}{\bibfnamefont{A.~S.} \bibnamefont{Ecker}},
  \bibinfo{author}{\bibfnamefont{P.}~\bibnamefont{Berens}},
  \bibinfo{author}{\bibfnamefont{A.~S.} \bibnamefont{Tolias}},
  \bibnamefont{and} \bibinfo{author}{\bibfnamefont{M.}~\bibnamefont{Bethge}},
  \bibinfo{journal}{Journal of Neuroscience} \textbf{\bibinfo{volume}{31}},
  \bibinfo{pages}{14272} (\bibinfo{year}{2011}).

\bibitem[{\citenamefont{Mejias and Longtin}(2014)}]{mejias2014}
\bibinfo{author}{\bibfnamefont{J.~F.} \bibnamefont{Mejias}} \bibnamefont{and}
  \bibinfo{author}{\bibfnamefont{A.}~\bibnamefont{Longtin}},
  \bibinfo{journal}{Frontiers in computational neuroscience}
  \textbf{\bibinfo{volume}{8}}, \bibinfo{pages}{107} (\bibinfo{year}{2014}).

\bibitem[{\citenamefont{Mejias and Longtin}(2012)}]{mejias2012Optimal}
\bibinfo{author}{\bibfnamefont{J.}~\bibnamefont{Mejias}} \bibnamefont{and}
  \bibinfo{author}{\bibfnamefont{A.}~\bibnamefont{Longtin}},
  \bibinfo{journal}{Physical Review Letters} \textbf{\bibinfo{volume}{108}},
  \bibinfo{pages}{228102} (\bibinfo{year}{2012}).

\end{thebibliography}

\end{document}